\pdfoutput=1
\RequirePackage{ifpdf}
\ifpdf
\documentclass[pdftex]{sigma}
\else
\documentclass{sigma}
\fi

\numberwithin{equation}{section}

\newtheorem{Theorem}{Theorem}[section]
\newtheorem*{Theorem*}{Theorem}
\newtheorem{Corollary}[Theorem]{Corollary}
\newtheorem{Lemma}[Theorem]{Lemma}
\newtheorem{Proposition}[Theorem]{Proposition}

\theoremstyle{definition}
\newtheorem{Definition}[Theorem]{Definition}

\newtheorem{Example}[Theorem]{Example}
\newtheorem{Remark}[Theorem]{Remark}

\begin{document}

\allowdisplaybreaks

\newcommand{\arXivNumber}{2511.22876}

\renewcommand{\PaperNumber}{067}

\FirstPageHeading

\ShortArticleName{Degenerate Addition Formulas of the KP Hierarchy and Applications}

\ArticleName{Degenerate Addition Formulas of the KP Hierarchy\\ and Applications}

\Author{Atsushi NAKAYASHIKI}

\AuthorNameForHeading{A. Nakayashiki}

\Address{Department of Mathematics, Tsuda University, 2-1-1, Tsuda-Machi, Kodaira, Tokyo, Japan}
\Email{\mail{atsushi@tsuda.ac.jp}}

\ArticleDates{Received December 09, 2025, in final form July 06, 2026; Published online July 16, 2026}

\Abstract{It is well known that tau functions of the KP hierarchy satisfy addition formulas. Among them we consider the formula which expresses a~tau function with shifted arguments by $2n$ parameters in terms of the same tau function with shifted arguments by two parameters in the form of determinant. We then take the limits of it tending some of parameters to zero. As an important special case we obtain the formula which connects the shifted tau function to the Wronskian of functions obtained by substituting parameter values into the spectral variable of the wave function. As an application, we prove the equivalence of vertex operators and Darboux transformations. As another application, we derive a~new addition formula for Riemann's theta functions of Riemann surfaces by considering theta function solutions of the KP hierarchy. It can be considered as a~limit of Fay's famous formula (43) in the book [\textit{Lecture Notes in Math.}, Vol.~352, Springer, Berlin, 1973]. But the formula we have derived is a~different limit from the limits of (43) considered in that book.}


\Classification{37K40; 37K10; 33E05; 14H70; 14H42}

\section{Introduction}
Motivated by the recent study by Kodama and others
\cite{CK2008,Kodama2017, KW2011,KW2014} on shapes of soliton solutions of the KP equation and their relations to totally non-negative Grassmannians, cluster algebras, etc. the author~\cite{Nak2024} has studied the action of vertex operators to quasi-periodic solutions of the KP hierarchy, that is, solutions corresponding to smooth algebraic curves~\cite{Krichever1977,Segal-Wilson1985}. The vertex operators of the KP hierarchy had been introduced
by Date et al.\ \cite{DJKM} and act on the space of solutions of the KP hierarchy. In~\cite{Nak2024}, it is shown that the solution obtained by applying a~vertex operator to a~solution corresponding to a~smooth algebraic curve $C$
becomes some solution corresponding to a~singular algebraic curve whose normalization is the curve $C$.
Moreover, the solution obtained in this way is shown to represent a~soliton on the quasi-periodic background, where the background wave is the solution corresponding to $C$ chosen at the beginning.

Darboux transformations of the KP hierarchy also carry solutions of the KP hierarchy to
themselves~\cite{H-van-de-Leur2001,Kakei,Matveev-Salle1991,Ovel1993}. In~\cite{Kakei}, Kakei has studied in detail Darboux transformations of
solutions corresponding to genus one algebraic curves motivated by the work of Li and Zhang~\cite{Li-Zhang2022} on elliptic solitons.

In~\cite{Nak2024}, graphs of solutions obtained by applying vertex operators
to genus one solutions are shown. Graphs of solutions obtained by applying Darboux transformations to solutions of genus one are shown in~\cite{Kakei}.
Surprisingly both graphs look quite similar.
So it is natural to expect that two transformations are equivalent.

In this paper, we show this using the addition formula of the KP hierarchy.
More precisely we first establish a~certain addition formula
 (see Theorem~\ref{main-th}) by taking a~limit of the known general addition
formula~\eqref{addition} proved in~\cite{Shigyo2013}.
Using this, we prove that the vertex operator and Darboux transformation, if the parameters of the
transformations are appropriately chosen, produce the same solution
from the same initial solution which is not necessarily a~solution corresponding to a~genus one curve. We will explain the results in more detail below.

The addition formula for tau functions of the KP hierarchy was discovered by M.~Sato and Y.~Sato~\cite{SS}. The simplest addition formula takes the form
\[
\alpha_{12}\alpha_{34}
\tau_{12}\tau_{34}
-
\alpha_{13}\alpha_{24}\tau_{13}\tau_{24}
+
\alpha_{14}\alpha_{23}\tau_{14}\tau_{23}
=0,
\]
where $\alpha_{ij}=\alpha_i-\alpha_j$,
\begin{gather*}
\tau_{ij}=\tau(t+[\alpha_i]+[\alpha_j]),
\qquad
t={}^t(t_1,t_2,t_3,\dots ),
\qquad
[\alpha]={}^t\left(\alpha,\frac{\alpha^2}{2},\frac{\alpha^3}{3},\dots \right).
\end{gather*}
In~\cite{Shigyo2013,Takasaki-Takebe1995}, this single equation is proved to be equivalent to the
KP hierarchy itself. From this formula, the addition formula of the determinant form~\eqref{addition} had been derived in~\cite{Shigyo2013}.
For mutually distinct parameters $\alpha_i$, $\beta_i$, $1\leq i\leq n$, it expresses
the shifted tau function $ \tau(t+\sum_{i=1}^n([\beta_i]-[\alpha_i]))$ in terms of $\tau(t+[\beta_i]-[\alpha_j])$, $1\leq i$, $j\leq n$, as a~certain determinant.

In order to prove the equivalence between vertex operators and Darboux
transformations, we need some formula for
$ \tau\bigl(t-\sum_{i=1}^n[\alpha_i]\bigr)$.
If we set $\beta_j=0$, $j=1,\dots ,n$, in~\eqref{addition}, we shall get a~desired formula.

The problem here is that the expression~\eqref{addition} contains
the difference product of $\{\beta_j\}$ in the denominator.
It means that we need to differentiate the determinant with respect to $\beta_j$ if we want to set
$\beta_j=0$. At a~glance such a~formula becomes very messy since if we rewrite the derivative
with respect to $\beta$ of the function like $\tau(t+[\beta])$ in terms of
 derivatives of $t_j$, $j\geq 1$,
derivatives with respect to all $t_j$ appear.

We have found that, if we use linear differential equations satisfied
by the wave function (cf.~\eqref{wave-function}), the
$\beta_1,\dots ,\beta_n\to 0$ limit of
 \eqref{addition} can be written
entirely in terms of the wave function and its $t_1$-derivatives.
Finally, we obtain a~beautiful formula which connects $ \tau\bigl(t-\sum_{i=1}^n[\alpha_i]\bigr)$ to the Wronskian of the functions obtained by substituting parameter values into the spectral variable
of the wave function (see Theorem~\ref{main-th}).

In a~similar manner, it is possible to take
the partial limits as
$\beta_1,\dots ,\beta_k\to 0$ or $\alpha_1,\dots ,\alpha_k\to 0$ with $k\leq n$ of
the formula~\eqref{addition}.
The results are given by Theorems~\ref{main-th-2} and~\ref{main-th-3} which, together with Theorem~\ref{main-th}, constitute main results of this paper.

As an application of Theorem~\ref{main-th}, the equivalence of vertex operators and Darboux transformations is
proved based on the expressions of solutions derived in~\cite{Kakei,Nak2021,Nak2024}.

As another application of main results, we prove identities of theta functions.
Any identity for tau functions implies an identity for theta functions of Riemann surfaces if we consider solutions of the KP hierarchy expressed by theta functions.
For example, the addition formula~\eqref{addition} corresponds
to Fay's celebrated formula in~\cite[equation~(43)]{Fay1973} for
 Riemann's theta function $\theta(v)$ of a~Riemann surface ${\mathcal C}$ of genus $g$. It is described as follows. Taking a~symplectic basis of the homology group of $C$, let ${\rm d}v_i$, $i=1,\dots ,g$, be the normalized basis of holomorphic one forms,
$
\iota(p)=\int_{p_\infty}^{p}{\rm d}v$, $
{\rm d}v={}^t({\rm d}v_1,\dots ,{\rm d}v_g)
$
 the Abel--Jacobi map with a~base point $p_\infty\in {\mathcal C}$ and $E(p,q)$ the prime form.
Then for $n \geq 1$, mutually distinct points $p_i, q_i\in {\mathcal C}$, $i=1,\dots ,n$, and
 $c\in {\mathbb C}^g$ such that $\theta(c)\neq 0$ the following identity holds:
\begin{gather}
\theta\left(\sum_{i=1}^n\iota(p_i)-\sum_{i=1}^n \iota(q_i)+c\right)\theta(c)^{n-1}
\frac{\prod_{i<j}E(p_i,p_j)E(q_j,q_i)}{\prod_{i,j}E(p_i,q_j)}
\nonumber
\\
\qquad=\det\left(\frac{\theta(\iota(p_i)-\iota(q_j)+c)}{E(p_i,q_j)}\right)_{1\leq i,j\leq n}.
\label{intro-4}
\end{gather}

In Fay's book~\cite{Fay1973},
two kinds of limits of~\eqref{intro-4} are computed. The first one is the limit as~${c \to c'}$, where $c'$ is such that $\theta(c')=0$ and $c'$ is not a~singular point of the theta divisor $(\theta(v)=0)$.
The second one is the limit as all $q_i$ approach the same point in the first limit.
We derive the formula for theta functions
 corresponding to Theorem~\ref{main-th}, which is given by Theorem~\ref{addition-th}.
 It~corresponds to the limit of~\eqref{intro-4}, where all $q_j$ tend to the same point $p_\infty$ while keeping $\theta(c)\neq 0$. Therefore, formulas in Theorem~\ref{addition-th} are
 different from two limit formulas of Fay.
 In the case of genus one, the formulas of Theorem~\ref{addition-th} had been considered and
 proved in~\cite{Li-Zhang2022} with the aid of some formulas of elliptic functions.

 Finally, we would like to mention recent papers~\cite{Wang-Chen-Cheng2025,Zabrodin2025}
 where Darboux transformations are studied from the view point of free fermions.
They are closely related with the content of this paper.
 In particular, in~\cite{Wang-Chen-Cheng2025}, determinant formulas of some expectation values of free fermions are derived. Taking a~generating function of them and using the boson-fermion correspondence, they correspond to the formulas in Theorem~\ref{main-th-3}.
 They are derived using the determinantal structure of the Darboux transformations of~\cite{He-Li-Cheng2002}.
While Theorem~\ref{main-th-3} is proved in another direct and simple way as explained above.
Moreover, in our theorems it is important for applications that the formulas are
written in the form of addition formulas. While the fermionic formulas
 in~\cite{Wang-Chen-Cheng2025} do not apparently take the form of addition formulas.

The present paper is organized as follows.
In Section \ref{sec2} after a~brief review of the KP hierarchy and the addition formula
for tau functions, the main theorems are stated and proved.
The vertex operators and related formulas are summarized in Section \ref{sec3}.
Section \ref{sec4} provides a~brief review of the Darboux transformation and the necessary formulas.
The equality of tau functions obtained by vertex operators and those obtained by Darboux transformations is proved in Section~\ref{sec5}.
In~Section~\ref{sec6}, new addition formula for Riemann's theta function of Riemann surfaces
is given. In the case of genus one, the addition formulas rewritten in terms of Jacobi--Riemann theta functions and Weierstrass functions without prime form are also given.

\section{KP hierarchy and addition formula for tau function}\label{sec2}
In this section, we first review the KP hierarchy~\cite{DJKM, Sato1981, SS} and addition formulas for
tau functions~\cite{SS,Shigyo2013}.
Then the main results of the paper are presented and proved.

Let
\smash{$
L(t)=\partial+\sum_{j=2}^\infty u_j\partial^{1-j}$},
$
t={}^t(t_1,t_2,t_3,\dots )$,
$
x=t_1$,
$
\partial=\partial/\partial x
$
be a~monic microdifferential operator of order one without $\partial^0$ term.
By defining, for $m\geq 1$ and a~function $f$,
\begin{gather*}
\partial^{-m}f=\sum_{j=0}^\infty \binom{-m}{j}f^{(j)}\partial^{-m-j},
\qquad
f^{(j)}=\frac{\partial^j f}{\partial x^j},
\end{gather*}
the set of microdifferential operators \smash{$\big\{\sum_{j\leq n} a_j \partial^j \text{ for some $n$}\big\}$},
where $a_j$'s belong to some ring of functions on which $\partial$ acts, becomes a~ring.
The differential operator part of \smash{$P=\sum_j a_j \partial^j$} is defined by
\smash{$P_+=\sum_{j\geq 0} a_j \partial^j$}.

The KP hierarchy is the set of nonlinear differential equations for $\{u_j\}_{j=2}^\infty$
given by
\begin{gather}
\frac{\partial L(t)}{\partial t_m}=\left[B_m,L(t)\right],
\qquad
B_m=\left(L(t)^m\right)_{+},\qquad m=1,2,3,\dots,
\label{KP-hierarchy}
\end{gather}
where $[B_m, L(t)]=B_mL(t)-L(t)B_m$.
In particular, $u=u_2$ satisfies the KP equation
\[
3u_{t_2t_2}+(-4u_{t_3}+6uu_x+u_{xxx})_x=0.
\]
It describes shallow water waves~\cite{Kodama2017}, where $(x,t_2)$ is the space
variables and $t_3$ is the time variable.

For a~solution $L(t)$ of the KP hierarchy, there exists a~function $\Psi(t,z)$, called
the wave function of $L(t)$,
of the form
\begin{gather*}
\Psi(t,z)=\left(1+\sum_{j=1}^\infty w_j(t) z^j\right){\rm e}^{\eta(t,z^{-1})},
\qquad
\eta(t,z)=\sum_{j=1}^\infty t_j z^{j},
\end{gather*}
which satisfies
\begin{gather}
\frac{\partial \Psi(t,z)}{\partial t_m}=B_m \Psi(t,z),\qquad
L(t)\Psi(t,z)=z^{-1}\Psi(t,z).
\label{linear-eq}
\end{gather}
So the variable $z$ is the spectral parameter.
 A~tau function of $L(t)$ can be introduced as a~function~$\tau(t)$ which satisfies
\begin{gather}
\Psi(t,z)=\frac{\tau(t-[z])}{\tau(t)}{\rm e}^{\eta(t,z^{-1})},
\qquad
[z]={}^t\left(z,\frac{z^2}{2},\frac{z^3}{3},\dots \right).
\label{wave-function}
\end{gather}
For an $L(t)$, a~tau function exists and unique up to constant multiples~\cite{DJKM}.

The operator $L(t)$ can be recovered from a~tau function through $\Psi(t,z)$ and
the KP hierarchy can be expressed as a~system of differential equations for $\tau(t)$ as
\begin{gather}
\mathop{{\rm Res}}_{z=0} \tau(t-s-[z])\tau(t+s+[z]){\rm e}^{-2\eta(s,z^{-1})}\frac{{\rm d}z}{z^2}=0,
\label{bilinear}
\end{gather}
where $s={}^t(s_1,s_2,\dots )$ is a~set of parameters. Expanding~\eqref{bilinear} in $s_1,s_2,\dots $, we have a~set
of differential equations for $\tau(t)$ in the form of so called Hirota's bilinear equations.

The function $u_2(t)$ is recovered from $\tau(t)$ by
$
u_2(t)=2\partial^2\log \tau(t)$.
It follows from~\eqref{bilinear} that, if $\tau(t)$ satisfies~\eqref{bilinear},
so does \smash{$ {\rm e}^{c_0+\sum_{j=1}^\infty c_i t_i}\tau(t)$}
 for any constants $c_i$, $i=0,1,2,\dots $.
This transformation
\smash{$
\tau(t) \to {\rm e}^{c_0+\sum_{j=1}^\infty c_i t_i}\tau(t)
$}
 is called a~gauge transformation.

It is discovered in~\cite{SS} that a~tau function of the KP hierarchy
satisfies addition formulas and that the totality of addition formulas is equivalent to the KP hierarchy.

The following addition formula of the determinant form is proved in~\cite{Shigyo2013}.

\begin{Theorem}[\cite{Shigyo2013}]
Let $n\geq 1$ and $\alpha_1,\dots ,\alpha_n$, $\beta_1,\dots ,\beta_n$ be mutually distinct parameters.
Then the following equation holds:
\begin{gather}
\tau\left(t+\sum_{i=1}^n[\beta_i]-\sum_{i=1}^n[\alpha_i]\right)\nonumber\\
\qquad
=\tau(t)^{-n+1}
\frac{\prod_{i,j=1}^n(\beta_i-\alpha_j)}{\prod_{i<j}(\alpha_{ij}\beta_{ji})}
\det\left(\frac{\tau(t+[\beta_i]-[\alpha_j])}{\beta_i-\alpha_j}\right)_{1\leq i,,j\leq n},
\label{addition}
\end{gather}
where $\alpha_{ij}=\alpha_i-\alpha_j$, $\beta_{ij}=\beta_i-\beta_j$.
\end{Theorem}

Our main aim is to get the formula for $ \tau\bigl(t-\sum_{j=1}^n[\alpha_j]\bigr)$.
To this end, we should set $\beta_j=0$ for all $j$ in~\eqref{addition}.
Notice here that the right-hand side of~\eqref{addition} contains a~term $\prod_{i<j}\beta_{ji}$ in the denominator. So we consider the limit $\beta_1,\dots ,\beta_n \to 0$.
To take the limit, we need to take derivatives with respect to $\beta_i$.
If we expand $\tau(t+[\beta_i]-[\alpha_j])$
in $\beta_i$, its coefficients contain derivatives of $\tau(t)$ in all $t_j$.
However, we can prove that the final result contains derivatives with respect to only the variable $x$ ($=t_1$). The result is described in the following theorem which,
together with Theorems~\ref{main-th-2} and~\ref{main-th-3}, constitute
 main results of this paper.

\begin{Theorem}\label{main-th}
Let $n\geq 1$ and $\alpha_1,\dots ,\alpha_n$ be mutually distinct non-zero parameters. Then we have
\[
\tau\left(t-\sum_{i=1}^n[\alpha_i]\right)
=\frac{1}{\prod_{i<j}\bigl(\alpha_j^{-1}-\alpha_i^{-1}\bigr)}
\operatorname{Wr}(\Psi(t,\alpha_1),\dots ,\Psi(t,\alpha_n))\tau(t){\rm e}^{-\sum_{i=1}^n\eta(t,\alpha_i^{-1})},
\]
where
\begin{gather*}
\operatorname{Wr}(\Psi(t,\alpha_1),\dots ,\Psi(t,\alpha_n))=\det\bigl(\Psi^{(i-1)}(t,\alpha_j)\bigr)_{1\leq i,j\leq n},
\qquad
\Psi^{(i)}(t,\alpha)=\frac{\partial^i \Psi(t,\alpha)}{\partial x^i}
\end{gather*}
is the Wronskian of $\Psi(t,\alpha_1),\dots ,\Psi(t,\alpha_n)$ with respect to the variable $x=t_1$.
\end{Theorem}

\begin{proof}
Let us set
$
f(t,\alpha,\beta)=\frac{\tau(t+[\beta]-[\alpha])}{\beta-\alpha}$.
Assuming $|\beta|<|\alpha|$, we consider $f(t,\alpha,\beta)$ as a~series of $\beta$ and
we write it as
\begin{gather}
f(t,\alpha,\beta)=\sum_{j=0}^\infty f_j(t,\alpha)\beta^j.
\label{f-expansion}
\end{gather}
Set
\begin{gather}
F=\frac{1}{\prod_{i<j}\beta_{ji}}D,
\qquad
 D=\det\left(f(t,\alpha_j,\beta_i)\right)_{1\leq i,j\leq n}
\label{def-F}
\end{gather}
and
\begin{gather}
F_0=F, \qquad F_{k}=F_{k-1}|_{\beta_{k}=0}, \qquad k\geq 1,\nonumber
\\
{\mathcal B}_k=\prod_{k\leq i<j\leq n}\beta_{ji}, \qquad 1\leq k\leq n-1,
\qquad
{\mathcal B}_n={\mathcal B}_{n+1}=1.\label{def-Fk}
\end{gather}
Notice that each $F_k$ is well defined due to~\eqref{addition}.

\begin{Lemma}
For $1\leq k\leq n$,
\begin{gather}
F_k=\frac{D_k}{{\mathcal B}_{k+1}\prod_{j=k+1}^n\beta_j^k},\label{Fk}
\end{gather}
where $D_k$ is the following determinant:
\begin{gather*}
D_k=
\left|
\begin{matrix}
f_0(t,\alpha_1)&\cdots&f_0(t,\alpha_n)\\
\vdots&\qquad&\vdots\\
f_{k-1}(t,\alpha_1)&\cdots&f_{k-1}(t,\alpha_n)\\
f(t,\alpha_1,\beta_{k+1})&\cdots&f(t,\alpha_n,\beta_{k+1})\\
\vdots&\qquad&\vdots\\
f(t,\alpha_1,\beta_{n})&\cdots&f(t,\alpha_n,\beta_{n})\\
\end{matrix}
\right|,
\end{gather*}
and the empty product \smash{$\prod_{j=k+1}^n\beta_j^k$} for $k=n$ is understood
to be $1$.
In particular, \begin{gather}
F_n=\det\left(f_{i-1}(t,\alpha_j)\right)_{1\leq i,j\leq n}.
\label{det-expression-1}
\end{gather}
\end{Lemma}

\begin{proof}
Notice that
\smash{$
{\mathcal B}_k|_{\beta_{k=0}}={\mathcal B}_{k+1}\prod_{j=k+1}^n\beta_j$}.
We prove~\eqref{Fk} by induction on $k$.
For $k=1$, we have
\begin{align*}
F_1
&=
F|_{\beta_1=0}
=
\frac{1}{{\mathcal B}_{2}\prod_{j=2}^n\beta_j}D|_{\beta_1=0}=
\frac{1}{{\mathcal B}_{2}\prod_{j=2}^n\beta_j}
\begin{vmatrix}
f_0(t,\alpha_1)&\cdots&f_0(t,\alpha_n)\\
f(t,\alpha_1,\beta_2)&\cdots&f(t,\alpha_n,\beta_2)\\
\vdots&\vdots&\vdots\\
f(t,\alpha_1,\beta_n)&\cdots&f(t,\alpha_n,\beta_n)\\
\end{vmatrix}
\\
&=
\frac{D_1}{{\mathcal B}_{2}\prod_{j=2}^n\beta_j}.
\end{align*}
Therefore,~\eqref{Fk} is valid.

{\samepage
Suppose that~\eqref{Fk} holds for $k$. Then
\begin{align*}
F_{k+1}
&=
F_k|_{\beta_{k+1}=0}
=
\frac{D_k}{{\mathcal B}_{k+1}\prod_{j=k+1}^n \beta_j^k}\Big|_{\beta_{k+1}=0}
=
\frac{1}{{\mathcal B}_{k+1}|_{\beta_{k+1}=0}\prod_{j=k+2}^n \beta_j^k}\bigl(D_k\beta_{k+1}^{-k}\bigr)\Big|_{\beta_{k+1}=0}
\\
&=
\frac{1}{{\mathcal B}_{k+2}\prod_{j=k+2}^n \beta_j^{k+1}}
\bigl(D_k\beta_{k+1}^{-k}\bigr)\Big|_{\beta_{k+1}=0}.
\end{align*}}
Here, by expanding each element of $(k+1)$th row of $D_k$ in $\beta_{k+1}$ using
\eqref{f-expansion}, we get
\begin{gather*}
D_k\beta_{k+1}^{-k}
=
\sum_{i=0}^\infty
\left|
\begin{matrix}
f_0(t,\alpha_1)&\cdots&f_0(t,\alpha_n)\\
\vdots&\vdots&\vdots\\
f_{k-1}(t,\alpha_1)&\cdots&f_{k-1}(t,\alpha_n)\\
f_{i}(t,\alpha_1)&\cdots&f_{i}(t,\alpha_n)\\
f(t,\alpha_1,\beta_{k+2})&\cdots&f(t,\alpha_n,\beta_{k+2})\\
\vdots&\vdots&\vdots\\
f(t,\alpha_1,\beta_{n})&\cdots&f(t,\alpha_n,\beta_{n})\\
\end{matrix}
\right|\beta_{k+1}^{i-k}.
\end{gather*}
Notice that the coefficient of $\beta_{k+1}^{i-k}$ for $0\leq i\leq k-1$ becomes zero
due to a~property of determinants.
Thus
\smash{$
\bigl(D_k\beta_{k+1}^{-k}\bigr)|_{\beta_{k+1}=0}=D_{k+1}
$}
and
\[
 F_{k+1}=
\frac{D_{k+1}}{{\mathcal B}_{k+2}\prod_{j=k+2}^n\beta_j^{k+1}},
\]
which shows~\eqref{Fk} for $k+1$. Therefore, \eqref{Fk} is proved.
\end{proof}

Next let us calculate the determinant of the right-hand side of~\eqref{det-expression-1}.
By~\eqref{wave-function},
\begin{gather}
\Psi(t+[\beta],z)=\frac{\tau(t+[\beta]-[z])}{\tau(t+[\beta])}{\rm e}^{\eta(t+[\beta],z^{-1})}.
\label{wave-shift}
\end{gather}
Assuming $|\beta|<|z|$, we have
\begin{align*}
{\rm e}^{\eta(t+[\beta],z^{-1})}
=
{\rm e}^{\eta(t,z^{-1})+\eta([\beta],z^{-1})}
=
{\rm e}^{\eta(t,z^{-1})+\sum_{j=1}^\infty \frac{1}{j}(\frac{\beta}{z})^j}
=
{\rm e}^{\eta(t,z^{-1})-\log(1-\frac{\beta}{z})}
=
\frac{z{\rm e}^{\eta(t,z^{-1})}}{z-\beta}.
\end{align*}
Putting it into~\eqref{wave-shift} and replacing $z$ by $\alpha$, we obtain
\begin{gather*}
\Psi(t+[\beta],\alpha)=-\alpha \frac{\tau(t+[\beta]-[\alpha])}{(\beta-\alpha)\tau(t+[\beta])}
{\rm e}^{\eta(t,\alpha^{-1})}.
\end{gather*}
Therefore, \begin{gather}
f(t,\alpha,\beta)=-\alpha^{-1}\Psi(t+[\beta],\alpha)\tau(t+[\beta]){\rm e}^{-\eta(t,\alpha^{-1})}.
\label{expression-fta}
\end{gather}
Let us calculate $f_j(t,\alpha)$ by expanding the right-hand side of~\eqref{expression-fta}
 in $\beta$.
For this purpose, let
\begin{gather}
{\rm e}^{\eta(t,z)}=\sum_{k=0}^\infty p_k(t) z^k,
\qquad
\tilde{\partial}={}^t(\partial_1,\partial_2/2,\partial_3/3,\dots ),
 \qquad \partial_k=\frac{\partial}{\partial t_k}.\label{def-pk}
\end{gather}

Notice that the Schur function $p_k(t)$ is a~homogeneous polynomial of degree $k$ if we define
$\deg t_k=k$. For example,
$
p_0=1$,
$
p_1=t_1$,
\smash{$
p_2=t_2+\frac{t_1^2}{2}$},
\smash{$
p_3=t_3+t_1t_2+\frac{t_1^3}{6}$}.
We have
\begin{gather*}
\begin{split}
& \Psi(t+[\beta],\alpha)
=
{\rm e}^{\eta(\tilde{\partial},\beta)}\Psi(t,\alpha)=
\sum_{k=0}^\infty p_k\bigl(\tilde{\partial}\bigr)\Psi(t,\alpha)\beta^k,
\\
& \tau(t+[\beta])
=
{\rm e}^{\eta(\tilde{\partial},\beta)}\tau(t)=
\sum_{k=0}^\infty p_k\bigl(\tilde{\partial}\bigr)\tau(t)\beta^k.
\end{split}
\end{gather*}
Therefore, \begin{gather}
\Psi(t+[\beta],\alpha)\tau(t+[\beta])
=\sum_{m=0}^\infty \sum_{k+l=m}
\bigl(p_k\bigl(\tilde{\partial}\bigr)\Psi(t,\alpha) p_l\bigl(\tilde{\partial}\bigr)\tau(t)\bigr)\beta^m.
\label{expansion-product}
\end{gather}

\begin{Lemma}
The equation
\begin{gather}
p_k\bigl(\tilde{\partial}\bigr)\Psi(t,\alpha)=
\bigl( \partial^k+(\text{lower order}) \bigr) \Psi(t,\alpha)
\label{diffop-univariable}
\end{gather}
holds.
\end{Lemma}

\begin{proof}
In the left-hand side of~\eqref{diffop-univariable},
 we can replace $\partial_j$ in $p_k\bigl(\tilde{\partial}\bigr)$ by
$\partial^j+(\text{lower order})$ using~\eqref{linear-eq}

Replacing $t_k$ by $\partial^k/k$ in~\eqref{def-pk}, we have
\begin{gather*}
\sum_{k=0}^\infty p_k\bigl(\tilde{\partial}\bigr)|_{\partial_j=\partial^j}z^k
=
{\rm e}^{\sum_{k=1}^\infty \frac{(z\partial)^k}{k}}
=
{\rm e}^{-\log(1-z\partial)}
=
\frac{1}{1-z\partial}
=
\sum_{k=0}^\infty \partial^k z^k.
\end{gather*}
Therefore, $
p_k\bigl(\tilde{\partial}\bigr)|_{\partial_j=\partial^j}=\partial^k$.
Taking into account that $p_k(t)$ is a~homogeneous polynomial, we~have the assertion
of the lemma.
\end{proof}

Using~\eqref{expression-fta}, \eqref{expansion-product} and~\eqref{diffop-univariable}, we have
\begin{gather}
f_l(t,\alpha)
=
C(t,\alpha)
\sum_{k=0}^lA_{l,k}\partial^k\Psi(t,\alpha),
\qquad
C(t,\alpha)=(-\alpha)^{-1} {\rm e}^{-\eta(t,\alpha^{-1})},
\label{fl}
\end{gather}
where $A_{l,k}$ is independent of $\alpha$ and $A_{l,l}=\tau(t)$. Namely,
\begin{gather*}
f_0(t,\alpha)=C(t,\alpha)\tau(t)\Psi(t,\alpha),
\qquad
f_1(t,\alpha)=C(t,\alpha)(\tau(t)\partial+A_{1,0})\Psi(t,\alpha),
\\
f_2(t,\alpha)=C(t,\alpha)\big(\tau(t)\partial^2+A_{2,1}\partial+A_{2,0}\big)\Psi(t,\alpha),
\qquad
\dots.
\end{gather*}
Substituting them into the right-hand side of~\eqref{det-expression-1} and using
the properties of determinants, we~have
\begin{gather}
F_n=\det(f_{i-1}(t,\alpha_j))=\tau(t)^n
\left(\prod_{j=1}^nC(t,\alpha_j)\right)
\det\bigl(\Psi^{(i-1)}(t,\alpha_j)\bigr).
\label{expression-Fn}
\end{gather}

By~\eqref{addition}, \eqref{def-F}, \eqref{def-Fk}, we have
\begin{gather*}
\tau\left(t-\sum_{i=1}^n[\alpha_i]\right)
=
\tau(t)^{-(n-1)}\frac{\prod_{i=1}^n (-\alpha_i)^n}{\prod_{i<j}\alpha_{ij}}F_n.
\end{gather*}
Substitute~\eqref{expression-Fn} into this equation and get
\begin{gather*}
\tau\left(t-\sum_{i=1}^n[\alpha_i]\right)
=
\frac{\prod_{i=1}^{n}\alpha_i^{n-1}}{\prod_{i<j}\alpha_{ij}}
\operatorname{Wr}(\Psi(t,\alpha_1),\dots ,\Psi(t,\alpha_n))\tau(t)
{\rm e}^{-\sum_{i=1}^n\eta(t,\alpha_i^{-1})}.
\nonumber
\end{gather*}
Finally, the proof of the theorem is completed if we notice
\begin{gather*}
\frac{\prod_{i=1}^{n}\alpha_i^{n-1}}{\prod_{i<j}\alpha_{ij}}
=
\frac{1}{\prod_{i<j}\bigl(\alpha_j^{-1}-\alpha_i^{-1}\bigr)}.\tag*{\qed}
\end{gather*} \renewcommand{\qed}{}
\end{proof}

In a~similar way, it is possible to construct the formula for
 $\tau\bigl(t+\sum_{i=1}^n[\beta_i]\bigr)$ in terms of the
Wronskian of the adjoint wave functions. Let
\begin{gather*}
\Psi^\ast(t,z)=\frac{\tau(t+[z])}{\tau(t)}{\rm e}^{-\eta(t,z^{-1})}
\end{gather*}
be the adjoint wave function of $L(t)$. It satisfies
\begin{gather*}
\frac{\partial \Psi^\ast(t,z)}{\partial t_m}=-B_m^\ast \Psi^\ast(t,z),
\end{gather*}
where $B_m^\ast$ is the formal adjoint of $B_m$ \cite{DJKM}.

\begin{Theorem}\label{main-th-2}
Let $n\geq 1$ and $\beta_1,\dots ,\beta_n$ be mutually distinct non-zero parameters.
 Then we have
\begin{gather*}
\tau\left(t+\sum_{i=1}^n[\beta_i]\right)=\frac{1}{\prod_{i<j}\bigl(\beta_i^{-1}-\beta_j^{-1}\bigr)}
\operatorname{Wr}(\Psi^\ast(t,\beta_1),\dots ,\Psi^\ast(t,\beta_n))\tau(t){\rm e}^{\sum_{i=1}^n\eta(t,\beta_i^{-1})}.
\end{gather*}
\end{Theorem}

\begin{proof} It is possible to prove the theorem by taking the limit as $\alpha_j\to 0$, $j=1,\dots ,n$, of~\eqref{addition} in a~similar manner to the proof of Theorem~\ref{main-th}.
Here we give another proof based on Theorem~\ref{main-th}.

It is easily proved that if $\tau(t)$ is a~solution of~\eqref{bilinear} so is $\tau(-t)$.
We apply Theorem~\ref{main-th} to~${\tilde{\tau}(t)=\tau(-t)}$ and use
\begin{gather}
\tilde{\Psi}(t,z)=\Psi^\ast(-t,z),
\label{tilde-psi-trf}
\end{gather}
where $\tilde{\Psi}(t,z)$ is the wave function of $\tilde{\tau}(t)$.
Then we substitute $t$ by $-t$ and obtain Theorem~\ref{main-th-2}.\looseness=1
\end{proof}

More generally, there are formulas which contain Theorems~\ref{main-th} and \ref{main-th-2} as
special cases. We simply tend $\beta_1,\dots ,\beta_m$ $\to 0$ or $\alpha_1,\dots ,\alpha_m$ $\to 0$ for $m\leq n$ in~\eqref{addition}.

We set
\begin{gather}
\Psi(t,\alpha,\beta)
=
\frac{-\alpha}{\beta-\alpha}\frac{\tau(t+[\beta]-[\alpha])}{\tau(t)}
{\rm e}^{\eta(t,\alpha^{-1})},
\label{psi-beta-alpha}
\\
\Psi^\ast(t,\alpha,\beta)
=
\frac{\beta}{\beta-\alpha}\frac{\tau(t+[\beta]-[\alpha])}{\tau(t)}
{\rm e}^{-\eta(t,\beta^{-1})}.\nonumber
\end{gather}

\begin{Theorem}\label{main-th-3} Let $m$, $n$ be nonnegative integers such that $n\geq m$.
\begin{itemize}\itemsep=0pt
\item[$(i)$] For mutually distinct non-zero parameters $\alpha_1,\dots ,\alpha_n$, $\beta_1,\dots , \beta_{n-m}$, we have
\begin{gather*}
\tau\left(t+\sum_{i=1}^{n-m}[\beta_i]-\sum_{i=1}^n[\alpha_i]\right)\\
\qquad
=
C_{m,n}
{\rm e}^{-\sum_{j=1}^n\eta(t,\alpha_j^{-1})}
\tau(t)
\left|
\begin{matrix}
\Psi(t,\alpha_1)&\cdots&\Psi(t,\alpha_n)\\
\vdots&\cdots&\vdots\\
\Psi^{(m-1)}(t,\alpha_1)&\cdots&\Psi^{(m-1)}(t,\alpha_n)\\
\Psi(t,\alpha_1,\beta_1)&\cdots&\Psi(t,\alpha_n,\beta_1)\\
\vdots&\cdots&\vdots\\
\Psi(t,\alpha_1,\beta_{n-m})&\cdots&\Psi(t,\alpha_n,\beta_{n-m})\\
\end{matrix}
\right|,
\end{gather*}
where
\begin{gather*}
C_{m,n}=(-1)^{n(m-1)}\frac{\prod_{j=1}^n\alpha_j^{m-1}\prod_{i=1}^{n-m}\prod_{j=1}^n(\beta_i-\alpha_j)}
{\prod_{i<j}^n\alpha_{ij}\prod_{i<j}^{n-m}\beta_{ji}\prod_{i=1}^{n-m}\beta_i^m}.
\end{gather*}

\item[$(ii)$] For mutually distinct non-zero parameters $\alpha_1,\dots ,\alpha_{n-m}$, $\beta_1,\dots ,\beta_n$, we have
\begin{gather*}
\tau\left(t+\sum_{i=1}^{n}[\beta_i]-\sum_{i=1}^{n-m}[\alpha_i]\right)\\
\qquad
=
C_{m,n}^\ast
{\rm e}^{\sum_{j=1}^n\eta(t,\beta_j^{-1})}
\tau(t)
\left|
\begin{matrix}
\Psi^\ast(t,\beta_1)&\cdots&\Psi^\ast(t,\beta_n)\\
\vdots&\cdots&\vdots\\
\Psi^{\ast(m-1)}(t,\beta_1)&\cdots&\Psi^{\ast(m-1)}(t,\beta_n)\\
\Psi^\ast(t,\alpha_1,\beta_1)&\cdots&\Psi^\ast(t,\alpha_1,\beta_n)\\
\vdots&\cdots&\vdots\\
\Psi^\ast(t,\alpha_{n-m},\beta_1)&\cdots&\Psi^\ast(t,\alpha_{n-m},\beta_n)\\
\end{matrix}
\right|,
\end{gather*}
where
\begin{gather*}
C^\ast_{m,n}=
(-1)^{m(n-1)}
\frac{\prod_{j=1}^n\beta_j^{m-1}\prod_{i=1}^{n}\prod_{j=1}^{n-m}(\beta_i-\alpha_j)}
{\prod_{i<j}^{n-m}\alpha_{ij}\prod_{i<j}^{n}\beta_{ji}\prod_{i=1}^{n-m}\alpha_i^m}.
\end{gather*}
\end{itemize}
\end{Theorem}

\begin{proof}
(i) Substitute~\eqref{fl} to~\eqref{Fk} with $k=m$ and change the index of $\beta_j$ as
$(\beta_{m+1},\dots ,\beta_n)$ to $(\beta_1,\dots ,\beta_{n-m})$. Then we get (i) of Theorem~\ref{main-th-3} after a~simple calculation.

(ii) As in the proof of Theorem~\ref{main-th-2}, we should simply apply the formula (i)
to $\tilde{\tau}(t)=\tau(-t)$. Let us denote the functions~\eqref{psi-beta-alpha}
corresponding to $\tilde{\tau}(t)$ by $\tilde{\Psi}(t,\beta,\alpha)$. Then
\begin{gather}
\tilde{\Psi}(t,\alpha,\beta)=\Psi^\ast(-t,\beta,\alpha).
\label{tilde-psi-beta-alpha-trf}
\end{gather}
Using~\eqref{tilde-psi-trf} and~\eqref{tilde-psi-beta-alpha-trf} and substituting $t$ by $-t$, we
obtain (ii) of the theorem.
\end{proof}

\section{Vertex operator}\label{sec3}
The results of the previous section are used to prove the equivalence
between solutions of the KP hierarchy generated by vertex operators and
those generated by Darboux transformations.
In this section, we recall the general formulas of solutions generated by vertex operators~\cite{Nak2024}.

Let
\begin{gather*}
X(p,q)={\rm e}^{\eta(t,p)-\eta(t,q)} {\rm e}^{-\eta(\tilde{\partial},p^{-1})+\eta(\tilde{\partial},q^{-1})}
\end{gather*}
be the vertex operator of the KP hierarchy~\cite{DJKM}.
It acts on a~function $f(t)$ as
\begin{gather}
X(p,q)f(t)={\rm e}^{\eta(t,p)-\eta(t,q)}f\bigl(t-\big[p^{-1}\big]+\big[q^{-1}\big]\bigr).
\label{action-X(p,q)}
\end{gather}

The importance of the vertex operator is stated in the following theorem.

\begin{Theorem}[\cite{DJKM}]
If $\tau(t)$ is a~solution of the KP-hierarchy, so is ${\rm e}^{aX(p,q)}\tau(t)$ for any
$a\in {\mathbb C}$.
\end{Theorem}

We consider the composition of ${\rm e}^{aX(p,q)}$ with various values of $a$, $p$, $q$, and
apply it to $\tau(t)$. The result can be written quite explicitly in a~simple form.

The vertex operators satisfy the following commutation relations:
\begin{gather*}
X(p_1,q_1)X(p_2,q_2)=X(p_2,q_2)X(p_1,q_1)
\qquad \text{if $p_1\neq q_2$ and $q_1\neq p_2$,} \\ X(p_1,q_1)X(p_2,q_2)=0
 \qquad \text{if $p_1=p_2$ or $q_1=q_2$}.
\end{gather*}
It follows that two vertex operators commute for general values
of parameters and $X(p,q)^2=0$. Then{\samepage
\begin{gather}
{\rm e}^{aX(p,q)}=1+aX(p,q).
\label{exp-X(p,q)}
\end{gather}
Based on these facts, we consider the following general composition of vertex operators.}

Let $M$, $N$ be positive integers, $L=M+N$,
$p_1,\dots ,p_L$ mutually distinct non-zero complex parameters and
$(a_{i,j})$ an $M\times N$ complex matrix. For these data, we set
\begin{gather}
G={\rm e}^{\sum_{i=1}^M\sum_{j=1}^N a_{i,j} X(p_{N+i}^{-1},p_j^{-1})}.
\label{VO-G}
\end{gather}
Let $\tau(t)$ be a~solution of the KP hierarchy. We apply $G$ to $\tau(t)$.
In order to have a~formula similar to soliton solutions in~\cite{Kodama2017},
we first make a~certain shift to $\tau(t)$, then apply $G$ and make a~gauge transformation.
In this way, we define a~new tau function $\tilde{\tau}(t)$ by
\begin{gather}
\tilde{\tau}(t)=\Delta\bigl(p_1^{-1},\dots ,p_N^{-1}\bigr){\rm e}^{\sum_{j=1}^N\eta(t,p_j^{-1})}
G \tau\left(t-\sum_{j=1}^N[p_j]\right),
\label{tau-VO}
\end{gather}
where
$
\Delta(p_1,\dots ,p_n)=\prod_{1\leq i<j\leq n}(p_j-p_i)$.
By~\eqref{action-X(p,q)} and~\eqref{exp-X(p,q)}, $\tilde{\tau}(t)$ is expressed as
linear combination of shifts of $\tau(t)$.
To give an explicit formula for it, let us define the $L\times N$ matrix~${B=(b_{i,j})}$~by
\begin{gather}
 B=\left(
 \begin{matrix}
 1&\quad&\quad\\
 \quad&\ddots&\quad\\
 \quad&\quad&1\\
 b_{N+1,1}&\cdots&b_{N+1,N}\\
 \vdots&\quad&\vdots\\
 b_{N+M,1}&\cdots&b_{N+M,N}\\
 \end{matrix}
 \right),
 \label{matrix-B1}
 \end{gather}
 where
\begin{gather}
b_{i,j}=\delta_{i,j}, \qquad 1\leq i, j\leq N,
\nonumber
\\
b_{N+i,j}=a_{i,j}\prod_{m\neq j}^N\frac{p_j^{-1}-p_m^{-1}}{p_{N+i}^{-1}-p_m^{-1}},
 \qquad 1\leq i\leq M,\qquad 1\leq j\leq N.
\label{matrix-B2}
\end{gather}
Notice that $a_{i,j}$ is recovered from $b_{N+i,j}$ uniquely by~\eqref{matrix-B2}.
Therefore, if a~set of parameters~${(p_1,\dots ,p_L)}$ is fixed, the set of $M\times N$
matrices $(a_{i,j})$ and the set of $L\times N$ matrices $B$ of the form~\eqref{matrix-B1}
 correspond bijectively.

We use the following notation
\begin{gather*}
[L]=\{1,\dots ,L\},
\qquad
\binom{[L]}{N}=\{(i_1,\dots ,i_N) \mid 1\leq i_1<\cdots<i_N\leq L\},
\end{gather*}
and for $I=(i_1,\dots ,i_N)\in \binom{[L]}{N}$
\begin{gather*}
\Delta^{-}_I=\Delta\bigl(p_{i_1}^{-1},\dots ,p_{i_N}^{-1}\bigr),
\qquad
\eta_I^{-}=\sum_{i\in I} \eta\bigl(t,p_i^{-1}\bigr),
\qquad
[p_I]=\sum_{i\in I} [p_i],
\\
B_I=\det(b_{i_r,s})_{1\leq r,s\leq N}.
\end{gather*}

The formula for $\tilde{\tau}(t)$ is given in~\cite{Nak2021,Nak2024} as described
in the following proposition.

\begin{Proposition}[{\cite{Nak2021,Nak2024}}]\label{VO-solution}
The function $\tilde{\tau}(t)$ defined by~\eqref{tau-VO} is expressed as
\begin{gather}
\tilde{\tau}(t)=\sum_{I\in \binom{[L]}{N}} B_I \Delta^{-}_I {\rm e}^{\eta_I^{-}}\tau(t-[p_I]).
\label{VO}
\end{gather}
\end{Proposition}

This is the general formula for a~solution of the KP hierarchy generated
by vertex operators.

In~\cite{Nak2024}, it is shown that, if $\tau(t)$ is a~solution associated with a~nonsingular algebraic curve (Riemann surface), $\tilde{\tau}(t)$ is a~solution associated with a~singular algebraic curve whose normalization is the algebraic curve corresponding to $\tau(t)$.
In this case,
it is also observed in~\cite{Nak2024} that~$2\partial^2\log \tilde{\tau}(t)$ describes a~soliton on a~quasi-periodic background corresponding to $2\partial^2\log \tau(t)$.

\section{Darboux transformation}\label{sec4}
Similarly to vertex operators, Darboux transformation also takes a~solution of the KP hierarchy to another one.
In this section, we review the definition and properties of Darboux transformations of the KP hierarchy following~\cite{H-van-de-Leur2001, Kakei}. Our aim here is to present the general formula for the tau
function obtained by successive applications of elementary Darboux transformations.

We use the symbols $M$, $N$, $L$, $p_i$ as in Section~\ref{sec3}.
Let $L(t)$ be a~solution of the KP hierarchy.

\begin{Definition} A~function $\varphi(t)$ is called an eigenfunction of $L(t)$ if it satisfies
\[
\frac{\partial \varphi(t)}{\partial t_m}=B_m \varphi(t),
\qquad
B_m=(L(t)^m)_+.
\]
\end{Definition}

For an eigenfunction $\varphi(t)$ of $L(t)$, we set
\begin{gather*}
G_\varphi=\varphi\partial \varphi^{-1}=\partial-\partial (\log\varphi),
\qquad
L_\varphi(t)=G_\varphi L(t) G_{\varphi}^{-1},
\qquad
\tau_\varphi(t)=\varphi(t)\tau(t).
\end{gather*}

\begin{Theorem}[\cite{H-van-de-Leur2001,
Kac-van-de-Leur2018,Kakei,Ovel1993}]\label{Darboux-property}
Let $L(t)$ be a~solution of the KP hierarchy~\eqref{KP-hierarchy} and $\varphi(t)$ be an eigenfunction of $L(t)$.
Then
\begin{itemize}\itemsep=0pt
\item[$(1)$] The operator $L_\varphi(t)$ is a~solution of the KP hierarchy and $\tau_\varphi(t)$ is a~tau function of $L_\varphi(t)$.

\item[$(2)$] For any eigenfunction $\psi(t)$ of $L(t)$, $G_\varphi(\psi)$ is an eigenfunction of $L_\varphi(t)$.
\end{itemize}
\end{Theorem}

The transformation from $L(t)$ to $L_\varphi(t)$ or from $\tau(t)$ to $\tau_\varphi(t)$ is called an elementary Darboux transformation.

It is possible to repeat elementary Darboux transformations
using a~set of eigenfunctions of an initial solution $L(t)$ with the help of
this theorem

Let $N\geq 2$ and $\varphi_1,\dots , \varphi_N$ be eigenfunctions of $L(t)$.
We first consider the elementary Darboux transformation of $L(t)$ by
$\varphi_1$. By Theorem~\ref{Darboux-property}\,(2),
the functions
\begin{gather*}
G_{\varphi_1}(\varphi_j)=:\varphi^{[1]}_j,
\qquad
j=2,\dots ,N,
\end{gather*}
become eigenfunctions of $L_{\varphi_1}(t)$.
Notice that $G_\varphi(\varphi)=0$ for any $\varphi$. That is why
we consider~${j\geq 2}$ here.
Next we transform $L_{\varphi_1}(t)$ by \smash{$\varphi^{[1]}_2$}. Then
\begin{gather*}
G_{\varphi^{[1]}_2}\bigl(\varphi^{[1]}_j\bigr)=G_{\varphi^{[1]}_2}G_{\varphi_1}(\varphi_j)=:
\varphi^{[21]}_j,
\qquad
j=3,\dots ,N,
\end{gather*}
are eigenfunctions of
\begin{gather*}
(L_{\varphi_1})_{\varphi^{[1]}_2}(t)
=(G_{\varphi^{[1]}_2}G_{\varphi_1})L(t)(G_{\varphi^{[1]}_2}G_{\varphi_1})^{-1}.
\end{gather*}
We proceed in this way.
In general, for $i\geq 2$ and $j\geq i+1$, we set
\begin{gather*}
G_{i}=G_{\varphi_i^{[i-1,\dots ,1]}}\cdots G_{\varphi_3^{[21]}}
G_{\varphi_2^{[1]}}G_{\varphi_1},
\qquad
\varphi_j^{[i,\dots ,1]}= G_i(\varphi_j).
\end{gather*}
Then \smash{$\varphi_j^{[i,\dots ,1]}$} is an eigenfunction of $G_{i}L(t)G_{i}^{-1}$ by
 Theorem~\ref{Darboux-property}\,(2) and
\begin{gather*}
\varphi_i^{[i-1,\dots ,1]}\cdots \varphi_3^{[21]}\varphi_2^{[1]}\varphi_1\tau(t)
\end{gather*}
is a~tau function of $G_{i}L(t)G_{i}^{-1}$ by Theorem~\ref{Darboux-property}\,(1).
We set
\begin{gather}
\widehat{\tau}(t)=
\varphi_N^{[N-1,\dots ,1]}\cdots \varphi_3^{[21]}\varphi_2^{[1]}\varphi_1\tau(t).
\label{tau-hat}
\end{gather}

\begin{Example}For $N=2$, we have
\begin{gather*}
\widehat{\tau}(t)=\varphi_2^{[1]}(t)\varphi_1(t)\tau(t)=G_{\varphi_1}(\varphi_2)\varphi_1\tau(t)
=(\varphi_1\partial(\varphi_2)-\partial(\varphi_1)\varphi_2)\tau(t)=\operatorname{Wr}(\varphi_1,\varphi_2)\tau(t).
\end{gather*}
\end{Example}

The appearance of the Wronskian is a~general feature of the transformation.

\begin{Theorem}[\cite{H-van-de-Leur2001,Kac-van-de-Leur2018,Kakei,Ovel1993}] The equality
$\widehat{\tau}(t)=\operatorname{Wr}(\varphi_1,\dots ,\varphi_N)\tau(t)$ holds.
\end{Theorem}

Let $C=(c_{i,j})$ be an $L\times N$ complex matrix of rank $N$. We take
\begin{gather}
\varphi_j(t)=\sum_{i=1}^Lc_{i,j}\Psi(t,p_i),
\qquad
j=1,\dots ,N,
\label{varphij-special}
\end{gather}
as a~set of eigenfunctions of $L(t)$.
Then
\begin{gather*}
\operatorname{Wr}(\varphi_1,\dots ,\varphi_N)
=\det\bigl(\bigl(\Psi^{(i-1)}(t,p_j)\bigr)_{1\leq i\leq N,1\leq j\leq L} C\bigr).
\end{gather*}
Using the Binet--Cauchy formula, we have
\[
\operatorname{Wr}(\varphi_1,\dots ,\varphi_N)=\sum_{I=(i_1,\dots ,i_N)\in \binom{[L]}{N}}
C_I \operatorname{Wr}(\Psi(t,p_{i_1}),\dots ,\Psi(t,p_{i_N})),
\label{pre-Darboux}
\]
where, for $I=(i_1,\dots ,i_N)$,
$
C_I=\det(c_{i_r,s})_{1\leq r,s\leq N}$.

Thus we have the following proposition.

\begin{Proposition}\label{Darboux-composition} The tau function $\widehat{\tau}(t)$ defined by~\eqref{tau-hat} with
$\varphi_1,\dots ,\varphi_N$ given by~\eqref{varphij-special} is expressed as
\begin{gather}
\widehat{\tau}(t)=\tau(t)\sum_{I=(i_1,\dots ,i_N)\in \binom{[L]}{N}}
C_I \operatorname{Wr} (\Psi(t,p_{i_1}),\dots ,\Psi(t,p_{i_N}) ).
\label{Darboux}
\end{gather}
\end{Proposition}

This is the general formula for a~solution generated by Darboux transformations.
In the next section, we compare this formula with the formula~\eqref{VO} in
Proposition~\ref{VO-solution}.

\begin{Remark}\label{Grassmannian}
We denote by $\operatorname{Gr}(N,L)$ the Grassmannian which consists of $N$ dimensional
subspaces of ${\mathbb C}^L$. A~point of $\operatorname{Gr}(N,L)$ is represented by an
$L\times N$ matrix of rank $N$. Two~${L\times N}$ matrices $Y_1$ and $Y_2$ represent
the same point of $\operatorname{Gr}(N,L)$ if and only if $Y_1=Y_2G$ for some invertible $N\times N$
matrix $G$.

If the matrix $C$ in~\eqref{varphij-special} is replaced by $CG$ with an invertible constant matrix $G$, then~$\widehat{\tau}(t)$ changes to $\det(G)\widehat{\tau}(t)$ and
$u(t)=2\partial^2\log \widehat{\tau}(t)$ is unchanged. So if we are interested in
solutions~$u(t)$ of the KP equation, $C$ can be considered as an element
of $\operatorname{Gr}(N,L)$.
\end{Remark}

\section{Connecting vertex operators to Darboux transformation}\label{sec5}
In this section, we prove, using Theorem~\ref{main-th}, that two tau functions~\eqref{VO} and~\eqref{Darboux} obtained by
vertex operators and Darboux transformations respectively coincide.

Consider the term corresponding to
$ I=(i_1,\dots ,i_N)\in \binom{[L]}{N}$ in the formula~\eqref{VO}.
We take~${n=N}$, $(\alpha_1,\dots ,\alpha_N)=(p_{i_1},\dots ,p_{i_N})$ in Theorem~\ref{main-th}.
Then
\begin{gather*}
\Delta_I^{-}{\rm e}^{\eta_I^{-}}\tau(t-[p_I])=
 \operatorname{Wr}\left(\Psi(t,p_{i_1}),\dots ,\Psi(t,p_{i_N})\right)\tau(t).
 \end{gather*}

Substituting this formula into~\eqref{VO}, we have the following corollary.

\begin{Corollary}\label{VO-D-identity}
If two matrices $B$ of~\eqref{matrix-B1} and $C$ given just before~\eqref{varphij-special} coincide, the equation
\begin{gather}
\tilde{\tau}(t)=\widehat{\tau}(t)
\label{VO=Darboux}
\end{gather}
is valid.
\end{Corollary}

It looks that the matrix $B$ of~\eqref{matrix-B1} seems special compared with the matrix $C$ which is an~${L\times N}$ matrix of rank $N$.
But this is not the case.

Let $S_L$ be the set of permutations of $\{1,\dots ,L\}$.
By the reason explained in Remark~\ref{Grassmannian}, we~consider the matrix $C$ as an
element of $\mathrm{Gr}(N,L)$.
Then we can prove the following.

\begin{Proposition}\label{BC-equivalence}
For any $L\times N$ matrix $C$ from
$\operatorname{Gr}(N,L)$, there exists a~matrix $B$ and $\sigma\in S_L$ such that~\eqref{VO=Darboux} holds up to constant multiples, where $\tilde{\tau}(t)$ corresponds to
$\{B,(p_{\sigma(1)},\dots ,p_{\sigma(L)})\}$ and $\widehat{\tau}(t)$ corresponds to~${\{C,(p_1,\dots ,p_L)\}}$.
\end{Proposition}
The strategy of the proof is as follows.
We take $C$ in the reduced column
echelon form (cf.\ Example~\ref{example-gr24}).
We then change the order of rows of $C$
appropriately so that the resulting matrix~$\tilde{C}$ takes the form of~\eqref{matrix-B1} and define the components of $B$ in such a~way that $B=\tilde{C}$ holds.

\begin{Lemma}\label{BC-equivalence-2}
Let $C=(c_{i,j})$ be an arbitrary $L\times N$ matrix and $p_1,\dots ,p_L$ mutually distinct non-zero parameters.
For an element $\sigma$ of $S_L$, define the matrix $\tilde{C}=(\tilde{c}_{i,j})$ and parameter
$\tilde{p}_1,\dots ,\tilde{p}_L$
by
$
\tilde{c}_{i,j}=c_{\sigma(i),j}$,
$
\tilde{p}_i=p_{\sigma(i)}$,
Then
\begin{gather*}
\sum_{I\in \binom{[L]}{N}} C_I\operatorname{Wr}(\Psi(p_I))
=
\sum_{I\in \binom{[L]}{N}} \tilde{C}_I\operatorname{Wr}(\Psi(\tilde{p}_I)),
\end{gather*}
where, for $I=(i_1,\dots ,i_N)$, $C_I=\mathrm{det}(c_{i_r,s})_{1\leq r,s\leq N}$,
$\operatorname{Wr}(\Psi(p_I))=\operatorname{Wr}(\Psi(t,p_{i_1}),\dots ,\Psi(t,p_{i_N}))$ and similarly for
$\tilde{C}_I$,
$\operatorname{Wr}(\Psi(\tilde{p}_I))$ in which $c_{i_r,s}$ and $p_{i_j}$ are
 replaced by $\tilde{c}_{i_r,s}$ and $\tilde{p}_{i_j}$, respectively.
\end{Lemma}

\begin{proof}
For $I=(i_1,\dots ,i_N)\in\binom{[L]}{N}$ we have
\begin{align}
\tilde{C}_I\operatorname{Wr}(\Psi(\tilde{p}_I))
&=
\det(\tilde{c}_{i_r,s})_{1\leq r,s\leq N}
\det\bigl(\Psi^{(r-1)}(t,\tilde{p}_{i_s})\bigr)_{1\leq r,s\leq N}
\nonumber
\\
&=
\det(c_{\sigma(i_r),s})_{1\leq r,s\leq N}
\det\bigl(\Psi^{(r-1)}(t,p_{\sigma(i_s)})\bigr)_{1\leq r,s\leq N}.
\label{bi-delta-1}
\end{align}
Let
\begin{gather*}
\{\sigma(i_1),\dots ,\sigma(i_N)\}=\{l_1,\dots ,l_N\},
\qquad
l_1<\cdots<l_N,
\end{gather*}
 and $\tilde{I}=(l_1,\dots ,l_N)$.
Due to the skew symmetry of the determinant, we have
\[
\text{RHS of~\eqref{bi-delta-1}}
=
\det(c_{l_r,s})
\det\bigl(\Psi^{(r-1)}(t,{p}_{l_s})\bigr)
=
C_{\tilde{I}}
\operatorname{Wr}(\Psi(p_{\tilde{I}})),
\]
and
\begin{gather}
\tilde{C}_I\operatorname{Wr}(\Psi(p_I))=C_{\tilde{I}}
\operatorname{Wr}(\Psi(p_{\tilde{I}})).
\label{CIWI}
\end{gather}
The lemma follows from~\eqref{CIWI} by summing up in $I$,
since the map sending $(i_1,\dots ,i_N)$ to $(l_1,\dots ,l_N)$ is a~bijection from $\binom{[L]}{N}$ to itself.
\end{proof}

\begin{proof}[Proof of Proposition~\ref{BC-equivalence}]

Let $C\in \operatorname{Gr}(N,L)$ and $B$ a~matrix of the form~\eqref{matrix-B1} with
its components $b_{N+i,j}$ to be determined.
We can assume that $C$ is in reduced column echelon form
(see~\eqref{C-example}, for example).
We suppose that pivots of $C$ are in $(i_1,1),\dots ,(i_N,N)$ components with
$1\leq i_1<\cdots<i_N\leq L$.
Let $\{1,2,\dots ,L\}=\{i_1,\dots ,i_N,j_1,\dots ,j_{L-N})\}$ with $j_1<\cdots<j_{L-N}$.
Then we define $\sigma\in S_L$ by
\begin{gather*}
\sigma(i_1,\dots ,i_N,j_1,\dots ,j_{L-N})=(1,2,\dots ,L).
\end{gather*}
and the $L\times N$ matrix $\tilde{C}=(\tilde{c}_{i,j})$ by
$
\tilde{c}_{i,j}=c_{\sigma(i),j}$.
Then $\tilde{C}$ becomes the matrix of the form~\eqref{matrix-B1}.
We define $b_{N+i,j}$ by
$
b_{N+i,j}=\tilde{c}_{N+i,j}$.
Then the proposition follows from Lemma~\ref{BC-equivalence-2}.
\end{proof}

\begin{Example}\label{example-gr24} The case of $\operatorname{Gr}(2,4)$.
Consider the matrix $C$ given by
\begin{gather}
 C=\left(
 \begin{matrix}
 1&0\\
 a&0\\
 0&1\\
 b&c\\
 \end{matrix}
 \right).
 \label{C-example}
 \end{gather}
 and parameters $(p_1,p_2,p_3,p_4)$.
 We shall show how to construct the matrix
\[
 B=\left(
 \begin{matrix}
 1&0\\
 0&1\\
 b_{3,1}&b_{3,2}\\
 b_{4,1}&b_{4,2}\\
 \end{matrix}
 \right)
 \label{B-example}
 \]
 corresponding to $C$.
 The pivots of $C$ are in $(1,1)$, $(3,2)$ components. So $i_1=1$, $i_2=3$, $j_1=2$, $j_4=4$ and $\sigma\in S_4$ is defined by $\sigma(1,3,2,4)=(1,2,3,4)$, that is, $\sigma$ is the permutation of $2$ and~$3$. The matrix $\tilde{C}$ is obtained by exchanging
 2nd and 3rd rows of $C$,
 \[
 \tilde{C}=\left(
 \begin{matrix}
 1&0\\
 0&1\\
 a&0\\
 b&c\\
 \end{matrix}
 \right).
 \label{C-tilde-example}
 \]
 We define $b_{3,1}=a$, $b_{3,2}=0$, $b_{4,1}=b$, $b_{4,2}=c$. Then $B=\tilde{C}$.
 The associated parameters are $(p_{\sigma(1)},\dots ,p_{\sigma(4)})=(p_1,p_3,p_2,p_4)$.
 With this $B$ and parameters, $\tilde{\tau}(t)$ given by~\eqref{VO} and
 $\widehat{\tau}(t)$ given by~\eqref{Darboux} coincide.

 The vertex operator $G$ corresponding to $(C,(p_1,\dots ,p_4))$ is given by~\eqref{VO-G} with
 \begin{gather*}
 a_{1,1}=a\frac{p_2^{-1}-p_3^{-1}}{p_1^{-1}-p_3^{-1}},
 \qquad
 a_{1,2}=0,
\qquad
 a_{2,1}=b\frac{p_4^{-1}-p_3^{-1}}{p_1^{-1}-p_3^{-1}},
 \qquad
 a_{2,2}=c\frac{p_4^{-1}-p_1^{-1}}{p_3^{-1}-p_1^{-1}}.
 \end{gather*}
 \end{Example}

\section{New addition formula for theta functions of Riemann surfaces}\label{sec6}
If we consider a~solution of the KP hierarchy expressed by a~theta function of a~Riemann surface,
the addition formula of Theorem~\ref{main-th} for tau functions becomes that for a~theta function.
 In this section, we derive it.

Let ${\mathcal C}$ be a~compact Riemann surface of genus $g$, $\{A_i,B_i\}$ a~homology basis
with the intersections $A_i\cdot A_j=0$, $B_i\cdot B_j=0$, $A_i\cdot B_j=\delta_{i,j}$,
$\{{\rm d}v_i\}_{i=1}^g$ the normalized basis of Abelian differentials of the first kind, ${\rm d}v={}^t({\rm d}v_1,\dots ,{\rm d}v_g)$, \smash{$\Omega_{i,j}=\int_{B_j} {\rm d}v_i$}, $\Omega=(\Omega_{i,j})$ the normalized period matrix, $p_\infty$ a~point of ${\mathcal C}$, \smash{$\iota(p)=\int_{p_\infty}^p {\rm d}v$} the Abel map and $z$ a~local coordinate around $p_\infty$ that satisfies~${z(p_\infty)=0}$.
We notice that, as a~local coordinate around $p_\infty$, we always use the same local coordinate $z$ specified here unless otherwise stated.

Riemann's theta function with the characteristics $e=\binom{a}{b}$, $a,b\in {\mathbb R}^g$
is defined by
\begin{gather*}
\theta
\left[\begin{matrix}
a\\b
\end{matrix}\right]
(v,\Omega)=\sum_{m\in {\mathbb Z}^g}\exp\bigl(\pi{\rm i}{}^t(m+a) \Omega (m+a)+2\pi{\rm i} {}^t (m+a)(v+b)\bigr),
\end{gather*}
where $v={}^t(v_1,\dots ,v_g)$.
In the following, it is denoted by
$ \theta\left[\begin{smallmatrix}a\\b\end{smallmatrix}\right](v)$
by omitting the
$\Omega$ dependence, since we do not consider changing $\Omega$ in the
following.
Moreover, the theta function with the characteristics
$\left[\begin{smallmatrix}0\\0\end{smallmatrix}\right]$ is
denoted by $\theta(v)$.

We need the prime form $E(x,y)$, $x,y\in {\mathcal C}$ (see~\cite{Fay1973,KNTY1988,MumfordII} for details). It is defined as follows. Let $\epsilon=
\left[\begin{smallmatrix}
\epsilon'\\ \epsilon''
\end{smallmatrix}\right]$, $\epsilon',\epsilon''\in\frac{1}{2} {\mathbb Z}^g$ be a~non-singular odd half period, which means that $\epsilon$ satisfies the following
two conditions,
\[
\theta[\epsilon](-v)=-\theta[\epsilon](v),
\qquad
\frac{\partial \theta[\epsilon]}{\partial v_i}(0)\neq 0 \qquad \text{for some $i$}.
\]
It can be proved that such $\epsilon$ exists but it is not necessarily unique~\cite{Fay1973,MumfordII}.
We take any one of them.
Let $h_\epsilon(x)$ be such that
\[
 h_\epsilon(x)^2=\sum_{k=1}^g \frac{\partial \theta[\epsilon]}{\partial v_k}(0){\rm d}v_k(x).
 \]
 It can be considered as a~section of a~certain
holomorphic line bundle on ${\mathcal C}$ and is unique up to sign~\cite{Fay1973}.
We choose any sign and define
$h_\epsilon(x)$.
Then the prime form is defined by
\[
E(x,y)=\frac{\theta[\epsilon](\iota(y)-\iota(x))}{h_\epsilon(x)h_\epsilon(y)}.
\]
It is the section of a~certain holomorphic line bundle on ${\mathcal C}\times{\mathcal C}$ independent of $\epsilon$ and the choice of the sign of $h_\epsilon(x)$.
The prime form satisfies the following properties:
\begin{itemize}\itemsep=0pt
\item[(i)] $E(x,y)=-E(y,x)$,
\item[(ii)] $E(x,y)=0$ if and only if $x=y$,
\item[(iii)] for any point $p\in{\mathcal C}$, $x,y\in {\mathcal C}$ near $p$ and a~local coordinate $u$ around $p$ the expansion of~$E(x,y)$ in $u(y)$ at $u(x)$ is of the form
\begin{gather*}
E(x,y)\sqrt{{\rm d}u(x)}\sqrt{{\rm d}u(y)}=u(y)-u(x)+O\bigl((u(y)-u(x))^3\bigr).
\end{gather*}
\end{itemize}

For points $x$, $y$ around $p_\infty$,
using the local coordinates $z=z(x)$, $w=z(y)$,
 define $E(z,w)$ and $E(p_\infty,y)$ by
\begin{gather*}
E(x,y)=\frac{E(z,w)}{\sqrt{{\rm d}z}\sqrt{{\rm d}w}},
\qquad
E(p_\infty,y)=\left.E(x,y)\sqrt{{\rm d}z}\right|_{x=p_\infty}=\frac{E(0,w)}{\sqrt{{\rm d}w}}.
\end{gather*}
Notice that $E(p_\infty,y)$ depends on the choice of a~local coordinate $z$.

We define $v_{i,j}$, $q_{i,j}$ as the expansion coefficients with respect to $z=z(x)$, $w=z(y)$
\begin{gather}
{\rm d}v_i=\sum_{j=1}^\infty v_{i,j}z^{j-1}{\rm d}z \qquad \text{near $p_\infty$},
\label{vij}
\\
{\rm d}_z{\rm d}_w\log \frac{E(z,w)}{w-z}=\sum_{i,j=1}^\infty q_{i,j} z^{i-1}w^{j-1}{\rm d}z {\rm d}w
\qquad
\text{near $(0,0)$}.
\label{qij}
\end{gather}
Since $E(z,w)$ is skew symmetric in $z$ and $w$, we have $q_{i,j}=q_{j,i}$.

We set
\begin{gather}
{\mathcal V}=(v_{i,j})_{1\leq i\leq g, 1\leq j},
\qquad
{\mathcal V}_1={}^t(v_{1,1},v_{2,1},\dots ,v_{g,1}),
\label{V1}
\\
q(t|s)=\sum_{i,j=1}^\infty q_{i,j}t_i s_j,
\qquad
q(t)=q(t|t).
\label{qts}
\end{gather}

By Krichever's theory on Baker--Akhiezer function~\cite{Krichever1977}, solutions of the KP hierarchy
expressed by Riemann's theta function of Riemann surfaces had been derived~\cite{KNTY1988,Segal-Wilson1985}.

\begin{Theorem}[\cite{KNTY1988,Segal-Wilson1985}]
For any $c={}^t(c_1,\dots ,c_g) \in {\mathbb C}^g$,
\begin{gather}
\tau(t)={\rm e}^{\frac{1}{2} q(t)}\theta({\mathcal V}\cdot t+c)
\label{theta-sol}
\end{gather}
is a~solution of the KP hierarchy, where
\begin{gather*}
{\mathcal V}\cdot t+c={}^t\left(\sum_{i=1}^\infty v_{1,i}t_i+c_1,\dots ,\sum_{i=1}^\infty v_{g,i}t_i+c_g\right).
\end{gather*}
\end{Theorem}

We take~\eqref{theta-sol} as $\tau(t)$ in Theorem~\ref{main-th} and derive
the corresponding formula for $\theta(v)$.
To this end we need to prepare some notation.

Let ${\rm d}r_k$, $k=1,2,\dots $, be the differentials of the second kind with a~pole
only at $p_\infty$ normalized by the conditions
\begin{gather*}
\int_{A_i} {\rm d}r_k=0\qquad \text{for any $i$},
\qquad
{\rm d}r_k={\rm d}\left(\frac{1}{z^k}+O(1)\right) \qquad \text{near $p_\infty$}.
\end{gather*}
It is well known that such ${\rm d}r_k$ exists and unique if a~local coordinate around $p_\infty$ is fixed. The differential ${\rm d}r_k$ can be described using the prime form.
Let
\begin{gather*}
\omega(x,y)={\rm d}_x{\rm d}_y\log E(x,y)=\left( \frac{1}{(z-w)^2}+\sum_{i,j=1}^\infty q_{i,j}z^{i-1}w^{j-1}\right){\rm d}z{\rm d}w
\end{gather*}
be the bilinear meromorphic differential on ${\mathcal C}\times {\mathcal C}$ with a~double pole
on the diagonal $\{x=y\}$.
Due to the quasi-periodicity of the theta function, $\omega(x,y)$ satisfies
\begin{gather*}
\int_{A_i(x)}\omega(x,y)=\int_{A_i(y)}\omega(x,y)=0,
\qquad
1\leq i\leq g,
\end{gather*}
where \smash{$\int_{A_i(x)}$} signifies the integral over $A_i$ in $x$ variable and similarly
for \smash{$\int_{A_i(y)}$}.

We consider $w=z(y)$ as a~parameter.
Then we take the coefficient of ${\rm d}w$ of $\omega(x,y)$ and differentiate it several times in $w$.
In this way, we get the differential in $x$ variable which gives~${\rm d}r_k$. Explicitly,
\begin{gather}
{\rm d}r_k=-\frac{1}{(k-1)!}\frac{\partial^{k}}{\partial w^{k}}{\rm d}_z\log E(z,w)\vert_{w=0}
={\rm d}_z\left(\frac{1}{z^k}-\sum_{j=1}^\infty q_{j,k}\frac{z^j}{j}\right).
\label{drk-prime form}
\end{gather}
We denote $ \int^{x}{\rm d}r_k$ the indefinite integral of ${\rm d}r_k$ such that
the expansion near $p_\infty$ in $z=z(x)$ has no constant term, that is,
\[
\int^x {\rm d}r_k=\frac{1}{z^k}-\sum_{j=1}^\infty q_{j,k}\frac{z^j}{j}.
\]
In particular, \[
\int^x {\rm d}r_1=-\left.\frac{\partial}{\partial w}\log E(z,w)\right|_{w=0}.
\]
Consider points $p_1,\dots ,p_n$ of ${\mathcal C}$ and define, for $1\leq j\leq n$,
\begin{gather}
D_j
=\sum_{k=1}^g v_{k,1}\frac{\partial}{\partial v_k}+\int^{p_j} {\rm d}r_1.
\label{Dj-1}
\end{gather}
In terms of ${\rm d}v_k$ and $E(z,w)$, it can be written as
\begin{gather}
D_j=\sum_{k=1}^g \frac{{\rm d}v_k}{{\rm d}z}\Big|_{z=0}\frac{\partial}{\partial v_k}
-\left.\frac{\partial}{\partial w}\log E(z(p_j),w)\right|_{w=0}.
\label{Dj-2}
\end{gather}

With these notation, we can rewrite the formula in Theorem~\ref{main-th}
 in terms of $\theta(v)$ which is the main theorem in this section.

\begin{Theorem}\label{addition-th}
Let $n\geq 1$, $p_1,\dots ,p_n$ be mutually distinct points of ${\mathcal C}$ different
from $p_\infty$ and $c\in {\mathbb C}^g$ a~vector satisfying $\theta(c)\neq0$. Then we have the following assertions:
\begin{itemize}\itemsep=0pt
\item[$(1)$]
\begin{gather*}
\frac{\theta\bigl({\mathcal V}_1x+c-\sum_{j=1}^n \iota(p_j)\bigr)\prod_{i<j}E(p_j,p_i)}
{\theta({\mathcal V}_1x+c)\prod_{i=1}^n E(p_\infty,p_i)^{n-1}}
{\rm e}^{x\sum_{j=1}^n\int^{p_j}{\rm d}r_1}\\
\qquad
=\det\left(\partial^{i-1}\left(
\frac{\theta\left({\mathcal V}_1x+c-\iota(p_j)\right)}{\theta({\mathcal V}_1x+c)}
{\rm e}^{x\int^{p_j}{\rm d}r_1}\right)\right)_{1\leq i,j\leq n}.
\end{gather*}

\item[$(2)$]
\begin{gather*}
\frac{\theta\bigl(c-\sum_{j=1}^n \iota(p_j)\bigr)\prod_{i<j}E(p_j,p_i)}
{\theta(c)\prod_{i=1}^n E(p_\infty,p_i)^{n-1}}=
\det\left(
\phi_i(p_j)
\right)_{1\leq i,j\leq n},
\end{gather*}
where
\begin{gather*}
\phi_i(p_j)=\left.D_j^{i-1}\frac{\theta\left(v-\iota(p_j)\right)}{\theta(v)}\right|_{v=c},
\qquad
i\geq 1.
\end{gather*}
\end{itemize}
\end{Theorem}

\begin{Remark}
The formula (1), (2) of this theorem
can be considered as a~limit of
Fay's general addition formula~\eqref{intro-4}.
It is the limit as all $p_1,\dots ,p_n$ approach the same point $p_\infty$ while
keeping~$c$ so that $\theta(c)\neq 0$.
In~\cite{Fay1973}, two limits of~\eqref{intro-4} are computed.
The first one is the limit as $c$ tends to~$c'$ where $c'$ is a~non-singular point of
$(\theta(v)=0)$.
The second one is the limit, in the first limit, as all $p_j$ tend to the same point.
Therefore, the formulas (1), (2) are different from those in \cite[Corollary~2.19]{Fay1973}.
 Moreover, in~\cite{Fay1973} the notion of $x$ derivative is not introduced.
 The $x$ derivative structure in Theorem~\ref{addition-th}\,(2) is reflected in the fact
 that the coefficient of $\partial/\partial v_k$, $1\leq k\leq g$, in~$D_j$ consists of
 only the first term of the expansion of ${\rm d}v_k$ at $p_\infty$ (see~\eqref{vij}, \eqref{Dj-1}, \eqref{Dj-2}).
 \end{Remark}

\begin{proof}[Proof of Theorem~\ref{addition-th}]
We use the following formulas:
\begin{gather}
{\rm e}^{q([z]|[w])}=\frac{zw}{w-z}\frac{E(z,w)}{E(0,z)E(0,w)} \qquad \text{for $z\neq w$},
\qquad
{\rm e}^{\frac{1}{2}q([z]|[z])}=\frac{z}{E(0,z)}
\label{use-1}
\\
q(t|[z])=\eta\bigl(t,z^{-1}\bigr)-\sum_{i=1}^\infty t_i\int^p {\rm d}r_i,
\label{use-2}
\\
{\mathcal V}[z]=\iota(p)=\int_{p_\infty}^p {\rm d}v,
\label{use-3}
\end{gather}
where $z=z(p)$ in~\eqref{use-2} and~\eqref{use-3}.
Those formulas can be verified using~\eqref{qij}, \eqref{V1}, \eqref{qts}, \eqref{drk-prime form}.

Firstly, let us calculate the left-hand side of the formula of Theorem~\ref{main-th} divided by
$\tau(t)$.
Set~${\alpha_i=z(p_i)}$, $p_i\in {\mathcal C}$. By~\eqref{use-3}, we have
\begin{gather*}
\tau\left(t-\sum_{i=1}^n[\alpha_i]\right)
={\rm e}^{\frac{1}{2}q(t-\sum_{i=1}^n[\alpha_i])}\theta\left({\mathcal V}\cdot t-\sum_{i=1}^n\iota(p_i)+c\right).
\end{gather*}
Then, using~\eqref{use-1}, \eqref{use-2}, we have
\begin{align}
\frac{\tau\bigl(t-\sum_{i=1}^n[\alpha_i]\bigr)}{\tau(t)}
={}&
\frac{\prod_{i=1}^n\alpha_i}{\prod_{i<j}\bigl(\alpha_i^{-1}-\alpha_j^{-1}\bigr)}
\frac{\prod_{i<j}E(\alpha_i,\alpha_j)}{\prod_{i=1}^nE(0,\alpha_i)^n}
{\rm e}^{-\sum_{i=1}^n\eta(t,\alpha_i^{-1})}
\nonumber
\\
&
\times
\frac{\theta\bigl({\mathcal V}\cdot t-\sum_{i=1}^n \iota(p_i)+c\bigr)}{\theta({\mathcal V}\cdot t+c)}
{\rm e}^{\sum_{i=1}^\infty t_i\sum_{j=1}^n \int^{p_j} {\rm d}r_i}.
\label{LHS}
\end{align}

Next, let us calculate the Wronskian part of the right-hand side of the formula
in Theorem~\ref{main-th}.

The wave function at $z=\alpha_j$ becomes
\[
\Psi(t,\alpha_j)=\frac{\alpha_j}{E(0,\alpha_j)}
\frac{\theta({\mathcal V}\cdot t+c-\iota(p_j))}{\theta({\mathcal V}\cdot t+c)}
{\rm e}^{\sum_{k=1}^\infty t_k\int^{p_j} {\rm d}r_k}.
\]
Then
\begin{gather}
\operatorname{Wr}(\Psi(t,\alpha_1),\dots ,\Psi(t,\alpha_n))\nonumber
\\
\qquad=\prod_{i=1}^n\frac{\alpha_i}{E(0,\alpha_i)}
\det\left(
\partial^{i-1}
\left(
\frac{\theta({\mathcal V}\cdot t+c-\iota(p_j))}{\theta({\mathcal V}\cdot t+c)}
{\rm e}^{x\int^{p_j} {\rm d}r_1}
\right)
\right)
{\rm e}^{\sum_{j=1}^n\sum_{k=2}^\infty t_k\int^{p_j} {\rm d}r_k}.
\label{RHS}
\end{gather}
By~\eqref{LHS} and~\eqref{RHS}, the formula of Theorem~\ref{main-th} becomes
\begin{gather}
\frac{\theta\bigl({\mathcal V}\cdot t-\sum_{i=1}^n \iota(p_i)+c\bigr)}{\theta({\mathcal V}\cdot t+c)}
{\rm e}^{x\sum_{j=1}^n\int^{p_j} {\rm d}r_1}
\nonumber
\\
\qquad=
\frac{\prod_{i=1}^n E(0,\alpha_i)^{n-1}}{\prod_{i<j} E(\alpha_j,\alpha_i)}
\det\left(
\partial^{i-1}
\left(
\frac{\theta({\mathcal V}\cdot t+c-\iota(p_j))}{\theta({\mathcal V}\cdot t+c)}
{\rm e}^{x\int^{p_j} {\rm d}r_1}
\right)
\right).
\label{addition-1}
\end{gather}

The first product term of the right-hand side can be written as
\begin{gather*}
\frac{\prod_{i=1}^n E(0,\alpha_i)^{n-1}}{\prod_{i<j} E(\alpha_j,\alpha_i)}
=
\frac{
\prod_{i=1}^n \frac{E(0,z(p_i))^{n-1}}{\sqrt{{\rm d}z(p_i)}^{n-1}}
}
{
\prod_{i<j} \frac{E(z(p_j),z(p_i))}{\sqrt{{\rm d}z(p_j)}\sqrt{{\rm d}z(p_i)}}
}
=
\frac{\prod_{i=1}^n E(p_\infty,p_i)^{n-1}}{\prod_{i<j} E(p_j,p_i)}.
\end{gather*}

Substituting this into~\eqref{addition-1} and setting $t_1=x$, $t_j=0$ for $j\geq 2$, we get (1) of the theorem.

The formula (2) of the theorem follows from (1) as follows.
If we set
\[
f(v)=\frac{\theta(v-\iota(p_j))}{\theta(v)},
\qquad
v={}^t(v_1,\dots ,v_g),
\]
then
\[
\partial ( f({\mathcal V}_1 x+c) )
=\sum_{k=1}^g v_{k,1}\left.\frac{\partial f(v)}{\partial v_{k}}\right|_{v={\mathcal V}_1 x+c}.
\]
Therefore, \begin{gather}
{\rm e}^{-x\int^{p_j} {\rm d}r_1}\partial^{i-1}
\bigl({\rm e}^{x\int^{p_j} {\rm d}r_1} f({\mathcal V}_1 x+c) \bigr)
\nonumber
\\
\qquad=
\bigl({\rm e}^{-x\int^{p_j} {\rm d}r_1}\cdot \partial \cdot {\rm e}^{x\int^{p_j} {\rm d}r_1}\bigr)^{i-1}
f({\mathcal V}_1 x+c)
\nonumber
\\
\qquad=
\left(\partial +\int^{p_j} {\rm d}r_1\right)^{i-1}f({\mathcal V}_1 x+c)
\nonumber
\\
\qquad=\left.\left(\left(\sum_{k=1}^g v_{k,1}\frac{\partial}{\partial v_{k}}+\int^{p_j} {\rm d}r_1\right)^{i-1}
f(v)\right)\right|_{v={\mathcal V}_1 x+c}
=\bigl(D_j^{i-1}f(v)\bigr)\bigr|_{v={\mathcal V}_1 x+c}.
\label{derivative transformation}
\end{gather}

Now, multiply the formula (1) of the theorem by
\smash{$ {\rm e}^{-x\sum_{j=1}^n\int^{p_j} {\rm d}r_1}$},
use~\eqref{derivative transformation} and set $x=0$. Then we get~(2).
\end{proof}

{\bf The case of $\boldsymbol{ g=1}$.}
In the $g=1$ case, a~non-singular odd half period used to define the prime form is explicitly
known so that the formulas in Theorem~\ref{addition-th} can be rewritten in terms of theta functions only without using the prime form. We shall give them here.

Let $\omega_1$, $\omega_2$ be complex numbers such tat $\Omega=\omega_2/\omega_1$ has
positive imaginary part, $\sigma(u)$, $\zeta(u)$, $\wp(u)$ Weierstrass sigma, zeta, $\wp$-functions corresponding to $2\omega_1$, $2\omega_2$ respectively
\cite{Whittaker-Watson1902}.
We denote by $X$ the Weierstrass cubic
$
y^2=4x^3-g_2x -g_3$,
which is parametrized as $(x,y)=(\wp(u),\wp'(u))$.
 The segments $0(2\omega_1)$, $0(2\omega_2)$
define closed curves on $X$ which we denote by $A$, $B$. Then $\{A,B\}$
is a~basis of the homology group of $X$ with the intersections $A\cdot A=B\cdot B=0$, $A\cdot B=1$. Let~${{\rm d}u={\rm d}x/y}$. Then
\[
\int_A {\rm d}u=2\omega_1,
\qquad
\int_B{\rm d}u=2\omega_2,
\]
and the normalized differential of the first kind becomes
\smash{$
{\rm d}v=\frac{1}{2\omega_1}{\rm d}u$}.
The non-singular odd half period of $X$ is $\Omega/2+1/2$ whose characteristics
 is $\epsilon={}^t(1/2,1/2)$. We set
$
\theta_{11}(v)=\theta[\epsilon](v,\Omega)
$.
We take $p_\infty=\infty$ as a~base point and set
$
v(p)=\int_\infty^p {\rm d}v$,
which we adopt as a~local coordinate around $\infty$, $z=v(p)$.
Then $v_{1,j}=\delta_{1,j}$ and ${\mathcal V}_1={}^t(1,0,0,\dots )$.
We have
$
h_\epsilon^2(p)=\theta_{11}'(0){\rm d}v(p)$,
and the prime form is given by
\[
E(p,q)=\frac{\theta_{11}\bigl(\int_p^q {\rm d}v\bigr)}{h_\epsilon(p)h_\epsilon(q)}
=
\frac{\theta_{11}'(0)^{-1}\theta_{11}(z(q)-z(p))}{\sqrt{{\rm d}z(p)}\sqrt{{\rm d}z(q)}},
\]
where $\theta_{11}'(v)$ denotes the derivative of $\theta_{11}(v)$.
Therefore, $
E(z,w)=\theta_{11}'(0)^{-1}\theta_{11}(w-z)$.
We also have
\[
E(p_\infty,p)=\frac{E(0,z(p))}{\sqrt{{\rm d}z(p)}}=
\frac{\theta_{11}'(0)^{-1}\theta_{11}(z(p))}{\sqrt{{\rm d}z(p)}}.
\]
The normalized differential of the second kind ${\rm d}r_1$ is given by
\[
{\rm d}r_1=\frac{{\rm d}^2}{{\rm d}z^2}\log \theta_{11}(z) {\rm d}z.
\]
Set
$
\widehat{\zeta}(z)=\frac{{\rm d}}{{\rm d}z}\log\theta_{11}(z)$.
Then
$
D_j=\frac{\partial}{\partial v}+\widehat{\zeta}(\alpha_j)$,
$
\alpha_j=z(p_j)$.

{\bf Theorem~\ref{addition-th} for $\boldsymbol{ g=1}$: In terms of Jacobi--Riemann theta functions.}
Let $n\geq 1$, $\alpha_1,\dots ,\alpha_n \in \mathbb{C}$ be parameters which are mutually distinct and non-zero mod $\mathbb{Z}+\mathbb{Z}\Omega$ and $c\in \mathbb{C}$.
\begin{itemize}
\item[(1)] The following equation holds as a~function of $x$:
\begin{gather}
\frac{\theta\bigl(x+c-\sum_{j=1}^n \alpha_j\bigr)\prod_{i<j}\theta_{11}(\alpha_i-\alpha_j)}
{\theta(x+c)\prod_{i=1}^n\theta_{11}(\alpha_i)^{n-1}}
{\rm e}^{x\sum_{j=1}^n\widehat{\zeta}(\alpha_j)}\nonumber\\
\qquad=
\theta_{11}'(0)^{-\frac{1}{2}n(n-1)}
\det\left(\partial^{i-1}\left(\frac{\theta(x+c-\alpha_j)}{\theta(x+c)}
{\rm e}^{x\widehat{\zeta}(\alpha_j)}\right)\right)_{1\leq i,j\leq n}.
\label{addition:g=1}
\end{gather}

\item[(2)] If $\theta(c)\neq 0$, we have
\begin{gather*}
\frac{\theta\bigl(c-\sum_{j=1}^n \alpha_j\bigr)\prod_{i<j}\theta_{11}(\alpha_i-\alpha_j)}
{\theta(c)\prod_{i=1}^n\theta_{11}(\alpha_i)^{n-1}}
=
\theta_{11}'(0)^{-\frac{1}{2}n(n-1)}
\det\left(
\phi_i(\alpha_j)
\right)_{1\leq i,j\leq n},
\end{gather*}
where
\[
\phi_i(\alpha_j)=\left.D_{j}^{i-1}\left(\frac{\theta(v-\alpha_j)}{\theta(v)}\right)\right|_{v=c},
\qquad
i\geq 1.
\]
\end{itemize}

Next, we rewrite these formulas in terms of Weierstrass functions $\sigma(u)$, $\zeta(u)$.

Firstly, replace $c$ by $c+\Omega/2+1/2$ in~\eqref{addition:g=1}.
Then replace $x$, $c$, $\alpha_j$ by
$x/2\omega_1$, $c/2\omega_1$, $\alpha_j/2\omega_1$, respectively.
After that use
\begin{gather*}
\theta_{11}\left(\frac{u}{2\omega_1}\right)=
\frac{\theta_{11}'(0)}{2\omega_1}{\rm e}^{-\frac{\eta_1}{2\omega_1}u^2}\sigma(u),
\qquad
\eta_i=\zeta(\omega_i),
\\
\theta\left(v+\frac{\Omega}{2}+\frac{1}{2}\right)=-{\rm i}{\rm e}^{-\frac{\pi{\rm i}}{4}\Omega-\pi{\rm i} v}\theta_{11}(v).
\end{gather*}
We express the final formula using the Lam\'e function~\cite{Whittaker-Watson1902} (or Baker--Akhiezer function
of genus one~\cite{Krichever1977}) defined by
\[
\Phi(x,\alpha)=\frac{\sigma(x-\alpha)}{\sigma(x)\sigma(\alpha)}{\rm e}^{x\zeta(\alpha)}.
\]
Set
$
\Phi^{(i)}(x,\alpha)=\partial^i\Phi(x,\alpha)$.
Then we have the following formulas.

{\bf Theorem~\ref{addition-th} for $\boldsymbol{ g=1}$: In terms of Weierstrass functions.}
Let $n\geq 1$ and $\alpha_1,\dots ,\alpha_n$ be parameters which are mutually distinct and non-zero
mod $\mathbb{Z}(2\omega_1)+\mathbb{Z}(2\omega_2)$.
\begin{itemize}
\item[(1)] The following equation holds as a~function of $x$:
\begin{gather}
\frac{\sigma\bigl(x-\sum_{i=1}^n \alpha_j\bigr)\prod_{i<j}\sigma(\alpha_i-\alpha_j)}
{\sigma(x)\prod_{i=1}^n\sigma(\alpha_i)^n}
{\rm e}^{x\sum_{j=1}^n\zeta(\alpha_j)}
=
\det\bigl(\Phi^{(i-1)}(x,\alpha_j)\bigr)_{1\leq i,j\leq n}.
\label{g=1(1):Weierstrass}
\end{gather}

\item[(2)] If $c\in\mathbb{C}$ satisfies $\sigma(c)\neq 0$, we have
\[
\frac{\sigma\bigl(c-\sum_{i=1}^n \alpha_j\bigr)\prod_{i<j}\sigma(\alpha_i-\alpha_j)}
{\sigma(c)\prod_{i=1}^n\sigma(\alpha_i)^{n-1}}
=
\det\bigl(
\tilde{\phi}_{i}(\alpha_j)
\bigr)_{1\leq i,j\leq n},
\]
where
\begin{gather*}
\tilde{\phi}_{i}(\alpha_j)=\left(\frac{\partial}{\partial v}+\zeta(\alpha_j)\right)^{i-1}
\left.\left(\frac{\sigma(v-\alpha_j)}{\sigma(v)}\right)\right|_{v=c}.
\end{gather*}
\end{itemize}

\begin{Remark} Formula~\eqref{g=1(1):Weierstrass} is proved in \cite[Lemma 3.1]{Li-Zhang2022}
using the Frobenius--Stickelberger formula (see~\cite{Whittaker-Watson1902}, for example) and some related formulas proved in~\cite{DN2018,NSZ2019}.
\end{Remark}

\subsection*{Acknowledgements}

The author would like to thank Kanehisa Takasaki for valuable comments on the manuscript and for informing him the reference~\cite{Zabrodin2025}. He would also like to thank Yasuhiko Yamada for the comments on the possibility of the extension of Theorems~\ref{main-th} and~\ref{main-th-2}, which is now Theorem~\ref{main-th-3}, to Jipeng Cheng for
sending him papers~\cite{He-Li-Cheng2002,Wang-Chen-Cheng2025} and
to Xing Li for comments which helped him to revise the draft.
Finally, he thanks anonymous referees for careful reading the manuscript and for their valuable comments and corrections which helped him to improve the manuscript.
This work was supported by JSPS KAKENHI Grant Number JP24K06787.

\pdfbookmark[1]{References}{ref}
\LastPageEnding


\begin{thebibliography}{99}
\footnotesize\itemsep=0pt

\bibitem{CK2008}
Chakravarty S., Kodama Y., Classification of the line-soliton solutions of
 {KPII},
 \href{https://doi.org/10.1088/1751-8113/41/27/275209}{\textit{J.~Phys.~A}}
 \textbf{41} (2008), 275209, 33~pages,
 \href{http://arxiv.org/abs/0710.1456}{arXiv:0710.1456}.

\bibitem{DJKM}
Date E., Kashiwara M., Jimbo M., Miwa T., Transformation groups for soliton
 equations, in Nonlinear {I}ntegrable {S}ystems~-- {C}lassical {T}heory and
 {Q}uantum {T}heory ({K}yoto, 1981), World Scientific Publishing, Singapore,
 1983, 39--119.

\bibitem{Fay1973}
Fay J.D., Theta functions on {R}iemann surfaces, \textit{Lecture Notes in
 Math.}, Vol.~352, \href{https://doi.org/10.1007/BFb0060090}{Springer},
 Berlin, 1973.

\bibitem{He-Li-Cheng2002}
He J., Li Y., Cheng Y., The determinant representation of the gauge
 transformation operators,
 \href{https://doi.org/10.1142/S0252959902000444}{\textit{Chinese Ann. Math.
 Ser.~B}} \textbf{23} (2002), 475--486,
 \href{http://arxiv.org/abs/0904.1868}{arXiv:0904.1868}.

\bibitem{H-van-de-Leur2001}
Helminck G.F., van~de Leur J.W., Geometric {B}\"acklund--{D}arboux
 transformations for the {KP} hierarchy,
 \href{https://doi.org/10.2977/prims/1145477327}{\textit{Publ. Res. Inst.
 Math. Sci.}} \textbf{37} (2001), 479--519,
 \href{http://arxiv.org/abs/solv-int/9806009}{arXiv:solv-int/9806009}.

\bibitem{Kac-van-de-Leur2018}
Kac V.G., van~de Leur J.W., Equivalence of formulations of the {MKP} hierarchy
 and its polynomial tau-functions,
 \href{https://doi.org/10.1007/s11537-018-1803-1}{\textit{Jpn.~J. Math.}}
 \textbf{13} (2018), 235--271,
 \href{http://arxiv.org/abs/1801.02845}{arXiv:1801.02845}.

\bibitem{Kakei}
Kakei S., Solutions to the {KP} hierarchy with an elliptic background,
 \href{http://arxiv.org/abs/2310.11679}{arXiv:2310.11679}.

\bibitem{KNTY1988}
Kawamoto N., Namikawa Y., Tsuchiya A., Yamada Y., Geometric realization of
 conformal field theory on {R}iemann surfaces,
 \href{https://doi.org/10.1007/BF01225258}{\textit{Comm. Math. Phys.}}
 \textbf{116} (1988), 247--308.

\bibitem{Kodama2017}
Kodama Y., K{P} solitons and the {G}rassmannians, \textit{SpringerBriefs Math.
 Phys.}, Vol.~22, \href{https://doi.org/10.1007/978-981-10-4094-8}{Springer},
 Singapore, 2017.

\bibitem{KW2011}
Kodama Y., Williams L., K{P} solitons, total positivity, and cluster algebras,
 \href{https://doi.org/10.1073/pnas.1102627108}{\textit{Proc. Natl. Acad. Sci.
 USA}} \textbf{108} (2011), 8984--8989,
 \href{http://arxiv.org/abs/1105.4170}{arXiv:1105.4170}.

\bibitem{KW2014}
Kodama Y., Williams L., K{P} solitons and total positivity for the
 {G}rassmannian,
 \href{https://doi.org/10.1007/s00222-014-0506-3}{\textit{Invent. Math.}}
 \textbf{198} (2014), 637--699,
 \href{http://arxiv.org/abs/1106.0023}{arXiv:1106.0023}.

\bibitem{Krichever1977}
Krichever I.M., Methods of algebraic geometry in the theory of nonlinear
 equations,
 \href{https://doi.org/10.1070/RM1977v032n06ABEH003862}{\textit{Russian Math.
 Surveys}} \textbf{32} (1977), 185--213.

\bibitem{Li-Zhang2022}
Li X., Zhang D.-J., Elliptic soliton solutions: {$\tau $} functions, vertex
 operators and bilinear identities,
 \href{https://doi.org/10.1007/s00332-022-09835-4}{\textit{J.~Nonlinear Sci.}}
 \textbf{32} (2022), 70, 53~pages,
 \href{http://arxiv.org/abs/2204.01240}{arXiv:2204.01240}.

\bibitem{Matveev-Salle1991}
Matveev V.B., Salle M.A., Darboux transformations and solitons, \textit{Springer Ser.
 Nonlinear Dyn.}, \href{https://doi.org/10.1007/978-3-662-00922-2}{Springer},
 Berlin, 1991.

\bibitem{MumfordII}
Mumford D., Tata lectures on theta.~{II}. {J}acobian theta functions and differential equations, \textit{Progr. Math.}, Vol.~43,
 \href{https://doi.org/10.1007/978-0-8176-4578-6}{Birkh\"auser}, Boston, MA,
 1984.

\bibitem{Nak2021}
Nakayashiki A., Tau functions of~{$(n, 1)$} curves and soliton solutions on
 nonzero constant backgrounds,
 \href{https://doi.org/10.1007/s11005-021-01411-3}{\textit{Lett. Math. Phys.}}
 \textbf{111} (2021), 85, 31~pages,
 \href{http://arxiv.org/abs/2011.10691}{arXiv:2011.10691}.

\bibitem{Nak2024}
Nakayashiki A., Vertex operators of the~{KP} hierarchy and singular algebraic
 curves, \href{https://doi.org/10.1007/s11005-024-01836-6}{\textit{Lett. Math.
 Phys.}} \textbf{114} (2024), 82, 36~pages,
 \href{http://arxiv.org/abs/2309.08850}{arXiv:2309.08850}.

\bibitem{DN2018}
Nijhoff F., Delice N., On elliptic {L}ax pairs and isomonodromic deformation
 systems for elliptic lattice equations, in Representation {T}heory, {S}pecial
 {F}unctions and {P}ainlev\'e {E}quations~-- {RIMS}~2015, \textit{Adv. Stud.
 Pure Math.}, Vol.~76, \href{https://doi.org/10.2969/aspm/07610487}{The
 Mathematical Society of Japan}, Tokyo, 2018, 487--525,
 \href{http://arxiv.org/abs/1605.00829}{arXiv:1605.00829}.

\bibitem{NSZ2019}
Nijhoff F.W., Sun Y.-Y., Zhang D.-J., Elliptic solutions of {B}oussinesq type
 lattice equations and the elliptic {$N$}th root of unity,
 \href{https://doi.org/10.1007/s00220-022-04567-8}{\textit{Comm. Math. Phys.}}
 \textbf{399} (2023), 599--650,
 \href{http://arxiv.org/abs/1909.02948}{arXiv:1909.02948}.

\bibitem{Ovel1993}
Oevel W., Darboux theorems and {W}ronskian formulas for integrable
 systems.~{I}. {C}onstrained~{KP} flows,
 \href{https://doi.org/10.1016/0378-4371(93)90174-3}{\textit{Phys.~A}}
 \textbf{195} (1993), 533--576.

\bibitem{Sato1981}
Sato M., Soliton equations as dynamical systems on infinite dimensional
 {G}rassmann manifolds, \textit{RIMS Kokyuroku} \textbf{439} (1981), 30--46.

\bibitem{SS}
Sato M., Sato Y., Soliton equations as dynamical systems on
 infinite-dimensional {G}rassmann manifold, in Nonlinear {P}artial
 {D}ifferential {E}quations in {A}pplied {S}cience ({T}okyo, 1982),
 \textit{North-Holland Math. Stud.}, Vol.~81, North-Holland, Amsterdam, 1983,
 259--271.

\bibitem{Segal-Wilson1985}
Segal G., Wilson G., Loop groups and equations of {K}d{V} type,
 \href{https://doi.org/10.1007/BF02698802}{\textit{Inst. Hautes \'{E}tudes
 Sci. Publ. Math.}} \textbf{61} (1985), 5--65.

\bibitem{Shigyo2013}
Shigyo Y., On addition formulae of {KP}, m{KP} and {BKP} hierarchies,
 \href{https://doi.org/10.3842/SIGMA.2013.035}{\textit{SIGMA}} \textbf{9}
 (2013), 035, 16~pages,
 \href{http://arxiv.org/abs/1212.1952}{arXiv:1212.1952}.

\bibitem{Takasaki-Takebe1995}
Takasaki K., Takebe T., Integrable hierarchies and dispersionless limit,
 \href{https://doi.org/10.1142/S0129055X9500030X}{\textit{Rev. Math. Phys.}}
 \textbf{7} (1995), 743--808,
 \href{http://arxiv.org/abs/hep-th/9405096}{arXiv:hep-th/9405096}.

\bibitem{Wang-Chen-Cheng2025}
Wang C., Chen M., Cheng J., Toda {D}arboux transformations and vacuum
 expectation values,
 \href{https://doi.org/10.1016/j.geomphys.2024.105399}{\textit{J.~Geom.
 Phys.}} \textbf{209} (2025), 105399, 14~pages,
 \href{http://arxiv.org/abs/2408.09457}{arXiv:2408.09457}.

\bibitem{Whittaker-Watson1902}
Whittaker E.T., Watson G.N., A~course of modern analysis, 4th ed.,
Cambridge University Press,
 Cambridge, 1927.

\bibitem{Zabrodin2025}
Zabrodin A., Revisiting {B}\"acklund--{D}arboux transformations for {KP} and
 {BKP} integrable hierarchies,
 \href{https://doi.org/10.1016/j.geomphys.2026.105859}{\textit{J.~Geom.
 Phys.}} \textbf{227} (2026), 105859, 40~pages,
 \href{http://arxiv.org/abs/2506.07208}{arXiv:2506.07208}.

\end{thebibliography}
\end{document}